\documentclass[a4paper,11pt]{article}
\pdfoutput=1 

\usepackage{jinstpub} 

\usepackage{subfigure}
\usepackage{hhline}

\newcommand{\STEREO}{\textsc{Stereo}}

\title{The \STEREO{} experiment}

\author[a] {N.~Allemandou}
\author[b] {, H.~Almaz\'{a}n}
\author[a] {, P.~del Amo Sanchez}
\author[c] {, L.~Bernard}
\author[c] {, C.~Bernard}
\author[d] {, A.~Blanchet}
\author[d] {, A.~Bonhomme}
\author[c] {, G.~Bosson}
\author[c] {, O.~Bourrion}
\author[c] {, J.~Bouvier}
\author[b] {, C.~Buck}
\author[e] {, V.~Caillot}
\author[c] {, M.~Chala}
\author[d] {, P.~Champion}
\author[d] {, P.~Charon}
\author[d,1] {, A.~Collin \note{Present address: CEA DEN-Service d'\'etude des r\'eacteurs et de math\'ematiques appliqu\'ees (SERMA) CEA, Universit\'e Paris-Saclay F-91191, Gif-sur-Yvette, France}}
\author[d] {, P.~Contrepois}
\author[d] {, G.~Coulloux}
\author[e] {,  B.~Desbri{\`e}res}
\author[a] {, G.~Deleglise}
\author[a] {, W.~El Kanawati}
\author[a] {, J.~Favier}
\author[e] {, S.~Fuard}
\author[a] {, I.~Gomes Monteiro}
\author[b] {, B.~Gramlich}
\author[b] {, J.~Haser}
\author[c] {, V.~Helaine}
\author[c] {, M.~Heusch}
\author[e] {, M.~Jentschel }
\author[e] {, F.~Kandzia}
\author[f] {, G.~Konrad}
\author[e] {, U.~K{\"o}ster}
\author[c] {, S.~Kox}
\author[d] {, C.~Lahonde-Hamdoun}
\author[c,2] {, J.~Lamblin \note{Corresponding authors.}}
\author[d] {, A.~Letourneau}
\author[d,2] {, D.~Lhuillier}  
\author[c] {, C.~Li}
\author[b] {, M.~Lindner}
\author[a,3] {, L.~Manzanillas \note{Present address: LAL, Univ Paris-Sud, CNRS/IN2P3, Universit\'e Paris-Saclay, Orsay, France}}
\author[d] {, T.~Materna}
\author[c] {, O.~M{\'e}plan}
\author[d,e] {, A.~Minotti}
\author[e] {, C.~Monon}
\author[c] {, F.~Montanet}
\author[d] {, F.~Nunio}
\author[a] {, F.~Peltier}
\author[d] {, Y.~Penichot}
\author[d] {, M.~Pequignot}
\author[a] {, H.~Pessard}
\author[d] {, Y.~Piret}
\author[d] {, G.~Prono}
\author[g] {, G.~Qu\'em\'ener}
\author[c] {, J.-S.~Real}
\author[b] {, C.~Roca} 
\author[c] {, T.~Salagnac} 
\author[a] {, V.~Sergeyeva}
\author[b] {, S.~Schoppmann}
\author[d] {, L.~Scola}
\author[c] {, J.-P.~Scordilis}
\author[e] {, T.~Soldner}
\author[c] {, A.~Stutz}
\author[c] {, D.~Tourres}
\author[c] {, C.~Vescovi}
\author[c,4] {, S.~Zsoldos \note{Present address: Queen Mary University of London, School of Physics and Astronomy, London, United Kingdom}}

\affiliation[a]{Univ. Grenoble Alpes, Univ. Savoie Mont Blanc, CNRS/IN2P3, LAPP, 74000 Annecy, France}
\affiliation[b]{Max-Planck-Institut f\"ur Kernphysik, Saupfercheckweg 1, 69117 Heidelberg, Germany }
\affiliation[c]{Univ. Grenoble Alpes, CNRS, Grenoble INP, LPSC-IN2P3, 38000 Grenoble, France}
\affiliation[d]{IRFU, CEA, Universit\'e Paris-Saclay, F-91191 Gif-sur-Yvette}
\affiliation[e]{Institut Laue Langevin, CS 20156, 38042 Grenoble Cedex 9, France }
\affiliation[f]{Stefan-Meyer Institute, Boltzmanngasse 3, 1090 Wien, Austria  and Atominstitut, TU Wien, 1020 Wien, Austria }
\affiliation[g]{Normandie Univ, ENSICAEN, UNICAEN, CNRS/IN2P3, LPC Caen, 14000 Caen, France}

\emailAdd{jacob.lamblin@lpsc.in2p3.fr,david.lhuillier@cea.fr}

\usepackage{lineno}

\abstract{The \STEREO{} experiment is a very short baseline reactor antineutrino experiment aiming at testing the hypothesis of light sterile neutrinos as an explanation of the deficit of the observed neutrino interaction rate with respect to the predicted rate, known as the Reactor Antineutrino Anomaly. The detector center is located 10 m away from the compact, highly $^{235}$U enriched core of the research nuclear reactor of the Institut Laue Langevin in Grenoble, France. This paper describes the \STEREO{} site, the detector components and associated shielding designed to suppress the external sources of background which were characterized on site. It reports the performances in terms of detector response and energy reconstruction.}


\begin{document}

\maketitle
\flushbottom

\section{Introduction}
\label{sec:intro}
In preparation for high precision measurements of the survival probability of reactor electron antineutrinos at kilometer baseline experiments, the prediction of the antineutrino spectra emitted by nuclear reactor cores was re-evaluated in the early 2010's \cite{Mueller, Huber}, leading to a few percent increase of the predicted flux. The expected interaction rate was updated as well using the most recent neutron lifetime measurements in the calculation of the inverse beta-decay cross section. As a result, a $6.5\%$ mean deficit between the detected and expected interaction rates was found which is known as the Reactor Antineutrino Anomaly (RAA) \cite{Mention}. One possible explanation of this deficit is the oscillation towards light sterile neutrinos including a mass eigenstate in the eV range, so that oscillation patterns average out to the observed deficit after a few tens of meters.

Recent data from the Double Chooz \cite{Abe:2014bwa,Schoppmann}, Daya Bay \cite{An:2017} and RENO \cite{RENO:2015ksa} experiments confirm with high precision the mean antineutrino rate detected at short baselines. Taking advantage of the very high statistics available, the Daya Bay collaboration has analyzed the rates as a function of the $^{239}$Pu fission fraction. This result favors the scenario where most of the deficit is due to $^{235}$U~\cite{DB_U_Pu}. However, the precision is not sufficient to rule out the possible existence of sterile neutrinos~\cite{Hayes:2017res}. Global analyses including all available data still favor an oscillation toward sterile neutrinos \cite{GlobalFit, Giunti:2017yid}. Regarding the search for sterile neutrinos, the NEOS collaboration has published an exclusion contour in the (sin$^22\theta_{ee} - \Delta m^2_{41}$) plane \cite{NEOS} by comparing their measured spectrum with the experimental neutrino spectrum published by the Daya Bay collaboration \cite{An:2017}. The DANSS experiment, which suppresses the dependence on reactor inputs by comparing spectra measured with the same detector at different baselines, has also published first results \cite{DANSS}. A combined oscillation analysis \cite{Gariazzo:2017} points to a sterile neutrino with mixing parameters $\Delta m^2_{41}=1.3$~eV$^2$ and sin$^22\theta_{ee}=0.049\pm0.023\,(2\sigma)$ although final systematic uncertainties of the DANSS measurement are not available.

In this context, the \STEREO{} experiment provides a powerful complementary measurement. Installed at 10 m from the compact core research reactor of the Institut Laue Langevin (ILL) in Grenoble (France), \STEREO{} detects a virtually pure $^{235}$U fission neutrino spectrum, with negligible evolution expected during fuel cycles.  In order to mitigate the dependence on an external shape reference, the detector is segmented in six identical cells to measure the energy spectrum at different distances from the compact core. An oscillation pattern in the relative comparison of these six neutrino spectra would be an unambiguous signature of a flavor oscillation at very short baseline. This goal implies a good energy resolution, a control of the background and systematic uncertainties kept at the percent level. After describing the ILL site and the main components of the detector, the characterization of the external sources of background and the subsequent shielding design are presented. Then, the detector performances related to the energy measurement which were achieved during the first phases of data taking are reported.

\section{The ILL Site}
\label{sec:ILLSite}

The ILL is an international research center located in Grenoble (France) built to provide intense beams of thermalized neutrons. It operates about 40 instruments for experiments in various fields such as condensed matter research, nuclear and particle physics, chemistry and biology.

\subsection{The research reactor}
\label{subsec:Reactor}

The neutron source is a high flux reactor using a 93\% $^{235}$U enriched fuel and a heavy water moderator. The maximum neutron flux inside the moderator is about $1.5\cdot10^{15}$ neutrons per cm$^2$ per second, which is one of the highest continuous fluxes worldwide.
The reactor core consists of a single compact fuel element. The fuel is arranged in a hollow cylinder with an external diameter of 40~cm and a height of 80~cm. Such a compact core is particularly suitable for very short baseline neutrino oscillation studies. However, the nominal reactor power of 58.3 MW$_{\rm th}$ is substantially lower than that of commercial nuclear power reactors. The reactor can operate at nominal power during 45 days. In practice, the reactor operates at a slightly lower power during cycles of about 50~days each, alternating with shutdown periods for exchanging the single fuel element and for maintenance operations. A neutrino experiment can take benefit from reactor-off periods for measuring the cosmic background. In general, there are 3-4 cycles per year.

\subsection{Characterization of the neutrino source}
\label{subsec:NeutrinoSource}

As in most reactors, the determination of the thermal power $P_{th}$ of the ILL core is based on the balance of enthalpy at the primary circuit, filled with heavy water. The heavy water flow is determined from the measurement of the pressure drop downstream from a diaphragm installed inside the primary loop. Sub-leading contributions of other technical loops circulating light or heavy water, as well as the power of the pumps have been subtracted to correct the mean value of $P_{th}$, with negligible contribution to the total uncertainty. The average thermal power during the first data taking period of \STEREO{}, defined as phase-I, is found to be 55.0 MW with a total relative uncertainty of 1.7\%.

Because the nuclear fuel is highly enriched in $^{235}$U, this isotope dominates the number of fissions occurring in the core. According to reactor simulations, the $^{239}$Pu isotope contributes, averaged over one reactor cycle, to only 0.5\% of the total fission rate. Considering that about 6 neutrinos are emitted per fission, the total number of emitted neutrinos at nominal reactor power is $\simeq 10^{19}$ per second whereof one third is above the detection energy threshold.

\subsection{The experiment hall}
\label{subsec:ExperimentHall}

\begin{figure}[h]
\centering
\includegraphics[width=0.90\textwidth]{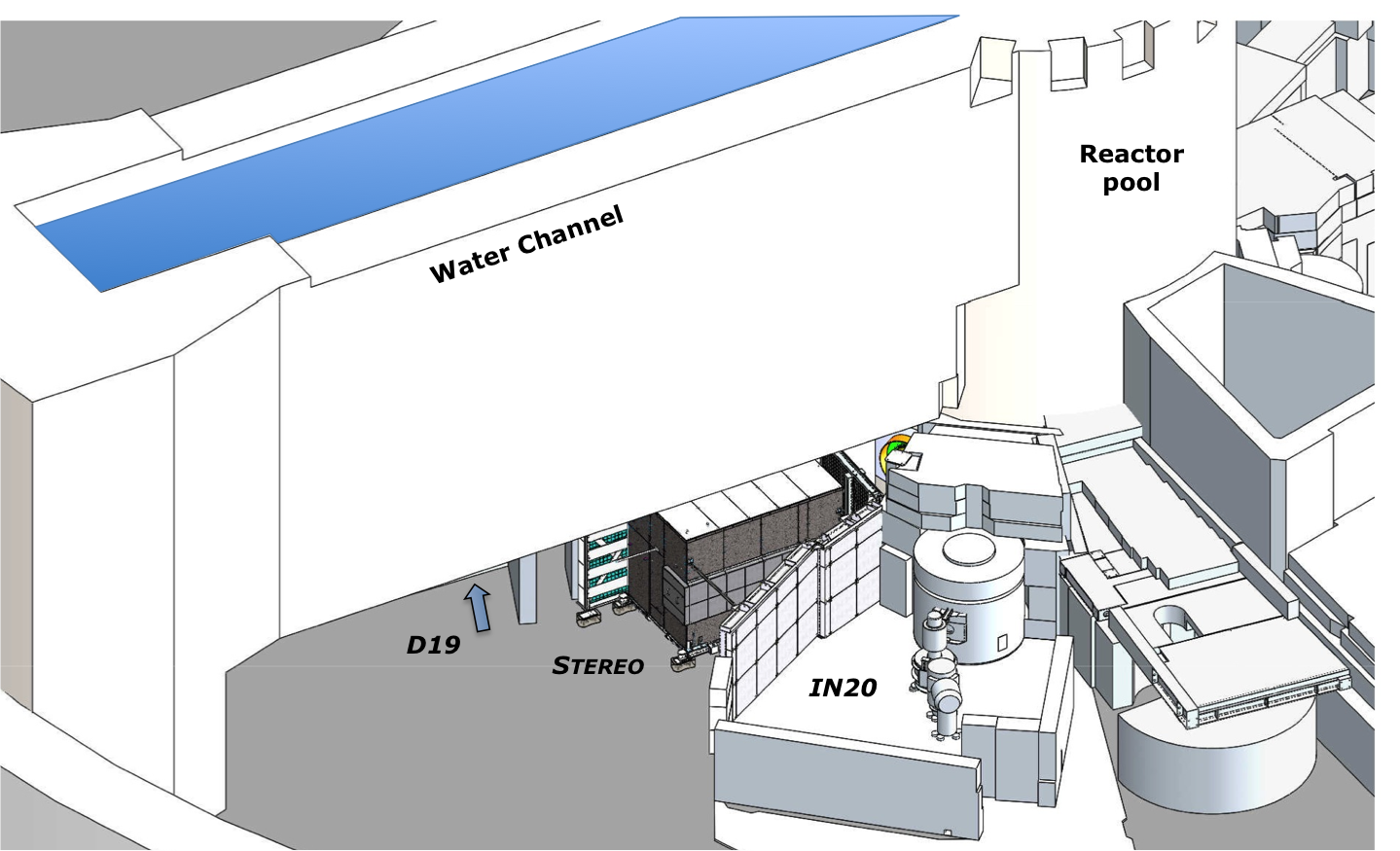}
\caption{3D view of the \STEREO{} site illustrating the water channel, the reactor pool and the neighboring experiment IN20. The D19 experiment is on the left of \STEREO{}, hidden by the water channel on this view.}
\label{fig:3DViewSite}
\end{figure}

The \STEREO{} detector, described in Section~\ref{sec:StereoDetector}, is located inside the reactor building in the ILL "level C" (see Figure~\ref{fig:3DViewSite}). In this experiment hall, intense neutron beams extracted from the reactor moderator are used to perform neutron scattering experiments. The two neighboring instruments of \STEREO{} are IN20 and D19. IN20 is a thermal three-axis spectrometer for inelastic scattering experiments in various sample environments. D19 is a thermal neutron diffractometer for single-crystal and fiber diffraction.

The vicinity of these instruments imposed several constraints on the \STEREO{} experiment.
Firstly, the instruments induce $\gamma$-rays and neutrons at the location of the \STEREO{} detector which need to be shielded.
Secondly, IN20 uses sample environments providing high magnetic fields, up to 15 Tesla, at the sample position. The stray fields at the \STEREO{} detector location can perturb the operation of the photomultiplier tubes, requiring  magnetic shielding.
Thirdly, detector and shielding designs had to be optimized to fit in the limited available space. Site background characterization, mitigation, and magnetic shielding are detailed in Section~\ref{sec:shieldings}. 
Finally, the \STEREO{} detector is located just below a water channel of 7~m depth, used to transfer and store spent fuel, which adds additional space constraints. The detector and the shielding had to be first mounted at another location accessible by crane and then moved to the final measuring position via air cushions. As a counterpart, the transfer channel offers additional shielding against cosmic background as discussed in Section~\ref{subsec:veto}.

\subsection{\STEREO{} run periods}
\label{subsec:RunPeriods}

After its installation in November 2016, the \STEREO{} detector recorded data over 2 reactor cycles, for a total of 84 days, and 2 reactor-off periods, for a total of 32 days \cite{StereoPhase1}.
This phase-I was followed by a long reactor shutdown of about 1 year devoted to reactor maintenance. The \STEREO{} detector had to be retracted in order to remove a beam tube that had reached the end of its lifetime. The \STEREO{} collaboration took advantage of this period to perform detector repairs and shielding improvements.
The phase-II of the experiment started in October 2017 during a reactor-off period and more than 120 days of reactor-off data could be acquired with a very smooth detector operation \cite{StereoPhase2Bg}. A 55-days reactor cycle started end of February 2018 and five more reactor cycles are planned until the end of 2019, allowing \STEREO{} to aim at more than 300 days of reactor-on data.

\section{The \STEREO{} Detector}
\label{sec:StereoDetector}

The \STEREO{} detector is filled with liquid scintillator (LS). A fraction of the energy deposited in the LS by charged particles is converted into light, which in turn is detected by photomultipliers. Neutral particles can generate a signal in the LS as well, induced by the recoils of charged particles. The dominant processes are electron recoils for $\gamma$-rays and proton recoils for fast neutrons. The \STEREO{} LS is a hydrogen-rich material serving both as a target and as an active detector for the interaction of the electron antineutrinos via the inverse beta-decay process (IBD):
\begin{equation}
  \label{eq:IBD}
  \bar{\nu}_e + p \rightarrow e^+ + n .
\end{equation}
A delayed coincidence may then be observed between a prompt positron event and a subsequent thermalized neutron capture event. The time delay between both events is used as a signature to discriminate antineutrinos from the majority of background events.
The incident neutrino spectrum is inferred from the measurement of the total positron signal (kinetic energy + annihilation $\gamma$-rays) since the positron's kinetic energy is closely related to the antineutrino energy~:
\begin{equation}
  \label{eq:Epositron}
  E_{e^+} = E_{\nu} - \Delta M + m_e = E_{\nu} - 0.782 \, \text{MeV},
\end{equation}
where $\Delta M $ is the mass difference between the neutron and the proton and $m_e$ is the electron mass. 

\subsection{Detector design}
 \label{subsec:DetectorDesign}

\begin{figure}[h]
\centering
\subfigure[]{\includegraphics[width=0.63\linewidth]{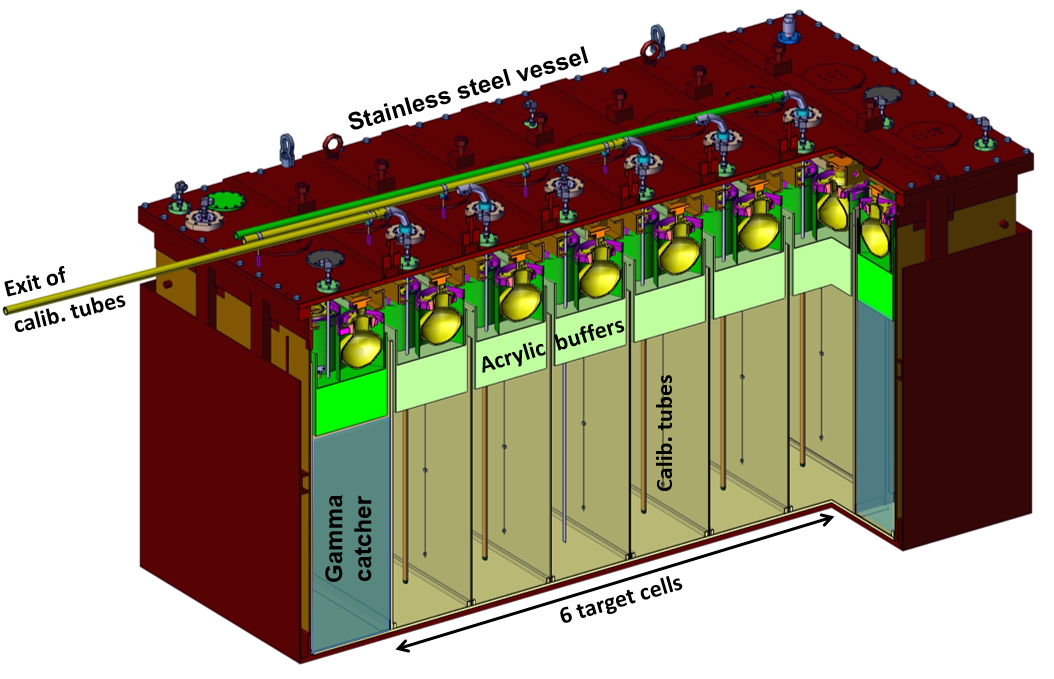}}\quad
\subfigure[]{\includegraphics[width=0.33\linewidth]{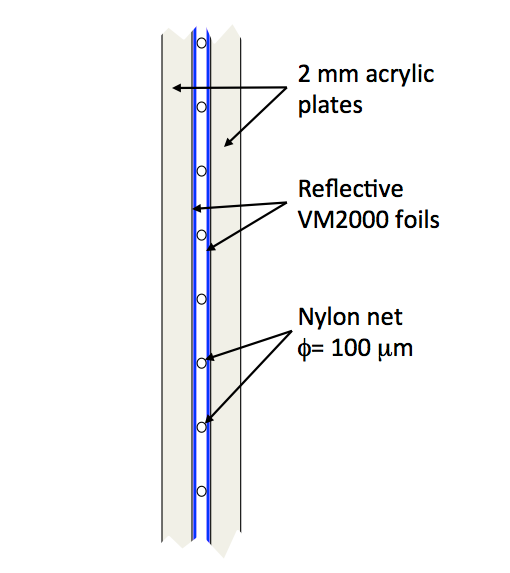}}\\
\caption{(a) Cut view of the \STEREO{} detector. (b) Schematic of the reflective wall structure.}
\label{fig:Det_cut_view}
\end{figure}

The innermost part of the detector (see Figure~\ref{fig:Det_cut_view}(a)) is the Target (TG). It consists of an acrylic aquarium with 12~mm thick walls of inner dimensions $L\times l \times h = 2.233\,\text{m} \times  0.889\,\text{m} \times 1.230\,\text{m} $. It is longitudinally divided in six identical and optically separated cells. This target volume is enclosed within a larger double walled stainless steel vessel (inner dimensions $L\times l \times h = 3.096\,\text{m} \times 1.536\,\text{m} \times 1.505\,\text{m}$), defining an outer crown around the central aquarium: the Gamma-Catcher (GC). This crown is designed to detect $\gamma$-rays escaping from events generated in the TG and it also serves as an active veto against external background entering the TG. The GC is divided into four cells~: two cells (GC front and GC back) on the sides of the first and the sixth TG cell, respectively, with the same geometry in order to suppress edge effects in the detector response, and two 30~cm thick cells (GC D19 and GC IN20) covering the full length at both sides (see Figure~\ref{fig:2DViewSite}).

The TG LS is doped with gadolinium in order to increase the cross-section of neutron capture reducing the capture time of IBD neutrons. The mean capture time in the \STEREO{} target has been measured to be 16\,$\mu$s, in agreement with simulations, allowing the use of a coincidence time window between prompt and delayed signal of below 70\,$\mu$s. From simulations, about 70\% of IBD neutrons are captured on gadolinium with variations of a few percent depending on the position in the detector. These values  Gadolinium allows also to take advantage of the high energy $\gamma$-rays following the capture. Details on the LS are given in Section \ref{subsec:LiquidScintillator}.

The scintillation light readout is done by 48 photomultiplier tubes (PMTs) located on top of the cells. These 8 inch PMTs, Hamamatsu R5912-100, have bialkali photocathodes with highest sensitivity in the wavelength region of the scintillation light emission, around 400~nm. The average quantum efficiency at this wavelength is about 35\%. The PMT bases have been optimized for low non-linearity, below 1\%, until up to 1500 photoelectrons (PE) per PMT. According to simulation, this corresponds to the maximum PMT signal for a 10~MeV neutrino interaction in the TG. The high-voltages have been adjusted to uniformize the PMTs' gain, the residual dispersion is below 1.5\%.
 
The PMTs are separated from the scintillator by thick acrylic blocks to achieve a more homogeneous light distribution among the PMTs of a cell and to mitigate the background from the PMTs' radioactive elements. There are 4 PMTs for each TG and short GC cell and 8 for each long GC cell. A bath of mineral oil above the acrylic blocks provides the optical coupling with the PMTs. To prevent potential absorption of water and oxygen by the LS, which would degrade its optical properties, a nitrogen atmosphere is maintained inside the detector vessel with a 10-30 mbar overpressure. All relevant parameters (temperatures, pressures, liquid levels, high-voltages, ...) are monitored by a slow control system.

All side walls and the bottom of every TG and GC cell are made highly reflective using a specular reflective film (VM2000$^{\text{TM}}$ Enhanced Specular Reflector (ESR)). The purpose of this reflective layer is threefold: to keep the cells optically separated, to collect a maximum amount of light in the PMTs and to reduce the position dependence in the collected light (top bottom effect). It is known that ESR films have a large reflectivity (above 98\%) for visible light in the air for almost all incident angles  \cite{Motta}. However, when the ESR films are immersed in a liquid, their reflectivity drops above a critical incident angle that depends on the light wavelength \cite{3M_private}. Thus, the reflective walls were designed to keep the reflection on the ESR film in air by placing the film in a sandwich between two acrylic plates and ensuring an air gap between the parts via a thin (100 $\mu$m) nylon net  (see Figure~\ref{fig:Det_cut_view}(b)). This way the walls act as perfect mirrors at large angles (total reflection on the air for angles above about 42$^\circ$) and as very good mirrors (reflectivity of VM2000 > 98\%) for smaller angles. It was not possible to build the entire surface of the wall reflective, especially at the borders of the walls where the sealing of the airgap allows for optical cross-talk between the cells. The detector was designed to keep the cross-talk between TG cells below 5\% and the cross-talk between the TG cells and the GC cells below 2\%. The total mass of the acrylic parts (aquarium and reflective walls) is about 220~kg.

In phase-I of the experiment, several deficiencies of the acrylics impacted the response of the inner detector. The most critical one was the loss of the oil-bath on top of two buffers -- one in the 4th TG cell, one in the GC front cell -- due to leaks of the buffer aquariums. This resulted in a drop of the collected light for the corresponding cells by a factor 2.5 compared to the other cells. Most acrylic sandwiches also lost their tightness at the level of their peripheral glue joint, allowing the LS to fill the air gap. To first order, this phenomenon did not affect the total light collection in the detector but the optical cross-talk between cells increased typically from 5 to about 15\%, making the response of the detector to energy deposits more complex to analyze. For this reason, a specific calibration and energy reconstruction procedure was developed, as presented in Sections \ref{subsec:cell-by-cell-calib} and \ref{sec:Erec}. The \STEREO{} detector was opened during the long reactor shutdown of 2017 and the leaking buffers and exchangeable separation plates were repaired for phase-II, which is now running with a symmetric detector at almost nominal light collection.

For an accurate normalization of the neutrino rates, surveys of the detector mechanics have been performed.  The dimensions of each empty cell were measured before filling. Once filled with LS, the liquid levels on both sides of the TG aquarium were monitored. All cells were found compatible with mean dimensions of $0.369\,\text{m} \times 0.892\,\text{m} \times 0.918\,\text{m}$. The corresponding volume is thus 0.3017~m$^3$, with a relative uncertainty of 0.9\% dominated by potential bending of the separation plates between the TG cells. The location and orientation of \STEREO{} with respect to the center of the reactor's fuel element was determined through successive surveys of the TG aquarium, the stainless steel vessel, the support structure, and reference points of the reactor building. Then, the already known position of the core center with respect to the reactor building could be used. At least two independent measurements were performed at each step of the survey and the associated uncertainties were estimated by taking the largest observed variations between measurements and/or expected values. The results are summarized in Table \ref{tab:survey}. The global uncertainty of 28 mm on the distance to the core is dominated by the uncertainty on the position of the core inside the reactor building (25 mm estimated from the survey of the symmetry axis of several neutron beam tubes). For the orientation of \STEREO{}, priority was given to the alignment of the detector with the water channel, hence the non-zero angle with respect to the direction of the core (see Figure~\ref{fig:3DViewSite}). From Table \ref{tab:survey}, the solid angle of the \STEREO{} TG is known at the 0.5\% level.

\begin{table}[h]
	\centering
	\begin{tabular}{| l || c |}
		\hline
		Distance Detector center - Core center & $10298 \pm 28$ mm\\
		Angle Detector axis - Direction to the core & $17.9 \pm 0.2$ deg\\
		\hline
	\end{tabular}
	\caption{Distance between the center of the \STEREO{} detector and the center of the reactor core and orientation of the detector axis with respect to the direction to the core.\label{tab:survey}}
\end{table}

\subsection{Liquid scintillator}
\label{subsec:LiquidScintillator}
The basic components of the \STEREO{} LS are the solvents, the wavelength shifting components (fluors) and, exclusively in the TG volume, the metalorganic gadolinium (Gd) complex. The LS solvent must fulfill the optical requirements of high transparency and light yield. Moreover, the liquid has to be compatible with the detector materials, such as the acrylic walls of the vessels. Safety aspects, especially a high flash point, are also very important since the experiment operates in the building of a nuclear reactor. All these specifications led to the use of linear alkyl benzene (LAB) as the principal component of the LS, corresponding to about 3/4 of the weight fraction in the liquids. Among the favorable features of LAB are its high transparency above 400~nm wavelength, a high flash point of $\sim$140$^\circ$C, and good compatibility with detector materials. To increase the light yield and pulse shape discrimination (PSD) capabilities,  ortho-phenyl-xylyl-ethane (PXE, $\sim$20~wt.\%) and di-isopropyl-naphtalene (DIN, $\sim$5~wt.\%) were added. In particular, the addition of DIN was found to significantly improve particle identification via PSD.

The Gd-loading is based on $\beta$-diketone chemistry. This approach of doping the rare earth element by dissolving Gd(thd)$_3$ molecules (Gd-tris-(2,2,6,6-tetramethyl-heptane-3,5-dionate) in the TG LS was first developed for Double Chooz \cite{DCLS} and later applied in the Nucifer experiment \cite{Nucifer}. The Gd-concentration in \STEREO{} of 0.2~wt.\% was optimized with respect to the coincidence time of the neutrino signal. A short coincidence time helps to lower the rather high accidental background rate in the ILL environment. The addition of Gd has only a small impact on the attenuation length of the LS, however the light yield is significantly quenched \cite{Aberle:2011zm} . A combination of the high quantum yield fluorescent molecules diphenyloxazole (PPO) as primary wavelength shifter and bis-methylstyrylbenzene (bis-MSB) as secondary wavelength shifter is utilized. The liquid scintillator of the GC is chemically identical except there is no Gd loading and there is less PPO. The PPO-concentration of the TG LS was increased from 3 to 7~g/l to partially compensate for the quenching effect of the Gd(thd)$_3$. The main basic properties of the \STEREO{} LSs are summarized in Table~\ref{tab:LSprop}.

\begin{table}

\begin{center}
\begin{tabular}{l|c|c|}

\cline{2-3}
 & Target (Gd-loaded) & GC (unloaded) \\ \hhline{~==}
\hhline{-~~}
\multicolumn{1}{ |c|| } {Attenuation length (430 nm)} & 6.9~m & 9.7~m\\
\multicolumn{1}{ |c|| } {Light yield (photons/MeV)} & 6500 & 8400 \\
\multicolumn{1}{ |c|| } {Density ($20^\circ$C)} & 0.887 & 0.884\\
\multicolumn{1}{ |c|| } {H-fraction (wt.\%)} & 11.45 & \\
\multicolumn{1}{ |c|| } {Gd-concentration (wt.\%)} & 0.2 & 0 \\
\multicolumn{1}{ |c|| } {Refractive index ($25^\circ$C)} & 1.50 & 1.50\\
\multicolumn{1}{ |c|| } {Flash point} & $74^\circ$C & $> 100^\circ$C\\ 
\hline

\end{tabular}
\caption[LS properties]{Basic properties of the \STEREO{} liquid scintillators from lab measurements.}
\label{tab:LSprop}
\end{center}
\end{table}

\begin{figure}[h]
	\centering
	\includegraphics[width=0.60\textwidth]{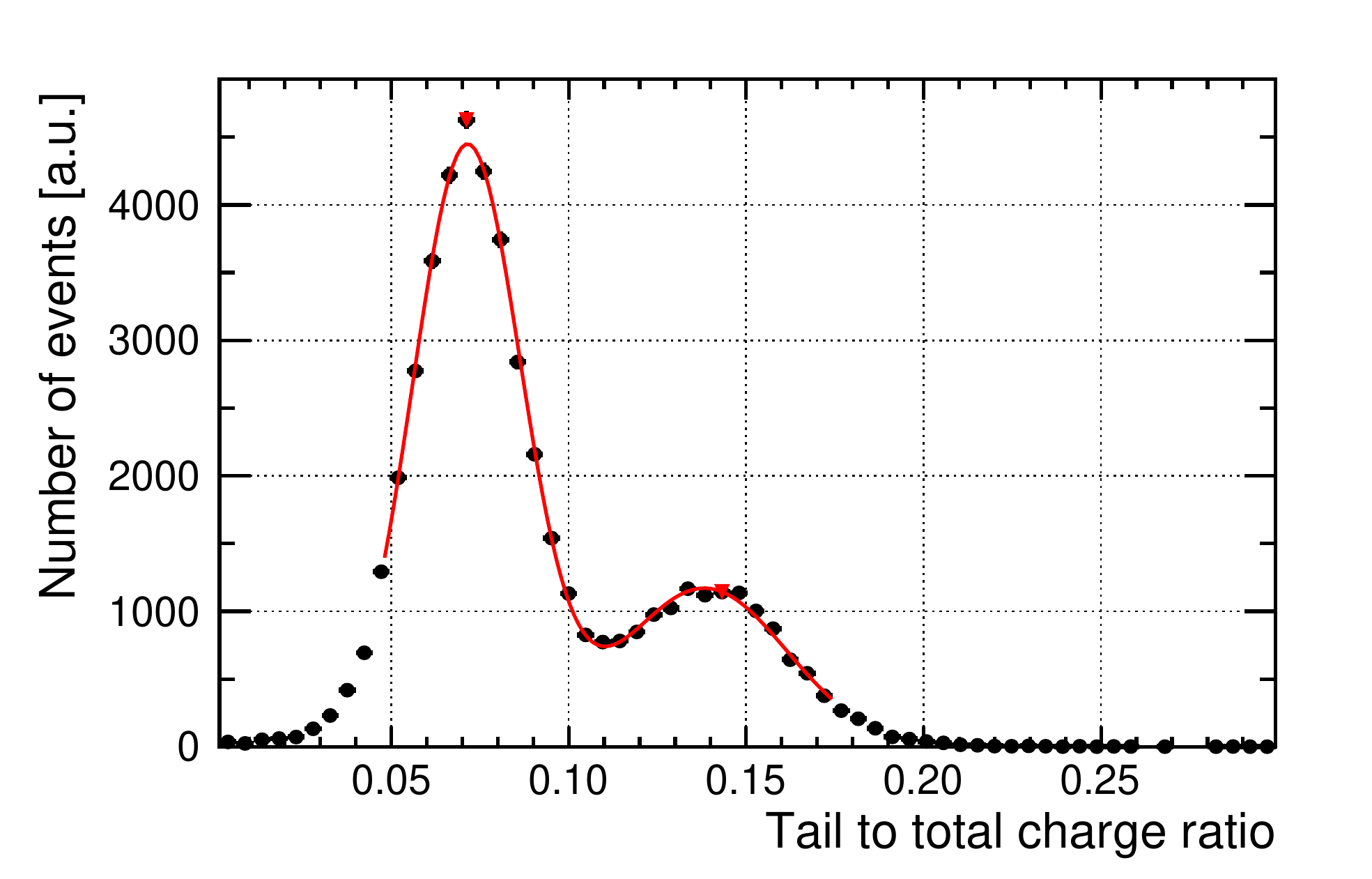}
	\caption{Pulse shape discrimination with the \STEREO{} detector at 2.2~MeV. The pulse tail charge to pulse total charge ratio is used to discriminate electron (left peak) and proton recoils (right peak) with a FoM of 0.71.}
	\label{fig:PSD}
\end{figure}

The PSD capability of the LS has been evaluated using an AmBe source and a small scintillator cell of 2 inch diameter. The PSD observable is calculated by dividing the pulse tail charge by the pulse total charge. The Factor of Merit (FoM) for discrimination between electron and proton recoils, defined as the difference between mean values of each population divided by the sum of their full width at half maximum, was found to be 0.87.  In the final detector, the FoM is lower due to the light propagation effects which enlarge the pulse width. At 2.2~MeV, corresponding to the hydrogen neutron capture peak, the FoM is 0.71. The corresponding distributions of electron and proton recoils is shown in Figure~\ref{fig:PSD}.

For the neutrino flux measurement, the proton number $n_p$ in the \STEREO{} target must be known with high accuracy. It mainly depends on the absolute liquid mass, $m$, and the relative hydrogen fraction, $f_H$, in the liquid ($n_p=\frac{f_H\cdot m}{m_H}$). The hydrogen mass, $m_H=1.673533\cdot10^{-27}$g, is known to very high accuracy and its uncertainty is neglected in the following. In total, $1602\pm2$~kg of liquid were used to fill the TG. The hydrogen fraction of a TG LS sample was determined at TU M\"unchen (TUM) by elemental analysis (CHN). The relative precision of this measurement is 1\%. A hydrogen fraction of $f_H=(11.45\pm0.11)$\% was obtained. With these numbers, the total proton number of the TG LS is $n_p=(1.096\pm0.011)\cdot10^{29}$.
 About 20 kg of the total mass are located in compensation volumes for thermal expansion, situated around the acrylic buffers, where the detection efficiency is lower. However, the differences of detection efficiency within the TG cells will be reproduced by the simulation.

In the phase-I of the experiment, one of the n-dodecane buffer volumes above the TG cells leaked into the TG volume. The mass of this spill was estimated to be $15\pm1$~kg with a hydrogen fraction of 15.3\% (calculated from molecular structure). This corresponds to an additional $1.37\cdot10^{27}$ protons, which adds 1.2\% to the protons in the TG liquid. 

\subsection{Calibration systems}
\label{subsec:calibration}

The calibration systems' purpose is to characterize all relevant aspects of the detector response. In the case of \STEREO{}, these include the liquid scintillator response to $\gamma$-rays and neutrons of different energies, the spatial dependence of light collection, PMT gains and the ability to reconstruct the neutron capture signal necessary for the IBD identification, as well as monitoring the change over time of the detector response.

PMT gains are calibrated every two hours via the single PE method \cite{bellamy}. The active volumes of the detector are illuminated with low-intensity light pulses generated by a 465~nm wavelength LED (see Section \ref{subsec:PMTCalibration}). The monitoring allows for detection of any short time scale variations due to PMT or electronics drift, and translates the charge seen by the PMTs into PEs.  In addition to the PMT gain calibration, LEDs are used to monitor the light attenuation of the liquid scintillator and to study the linearity of the PMTs and electronics over the required dynamic range by varying the intensity of the light pulses (see Section \ref{subsec:Linearity}). Another LED, emitting UV light, is used to directly excite the wavelength shifter.
 LEDs are placed within five identical boxes outside the shielding to illuminate all the TG and GC cells at three different heights, and the muon veto which is described in Section~\ref{subsec:veto}. Light is carried from the LED boxes through bundles of optical fibers to diffusive teflon balls placed in the detector liquid and used to ensure nearly isotropic light emission.

The remaining calibration observables are determined by deploying radioactive sources. The list of $\gamma$-ray and neutron calibration sources are shown in Table \ref{tab:set_sources}. These sources cover a range of $\gamma$-ray energies to anchor the energy scale and precisely study the non-linearity of the LS response. The sources $^{24}$Na, $^{65}$Zn and $^{124}$Sb were produced by neutron activation of natural NaCl, Zn and Sb, respectively, in a capture flux of about $2\cdot10^{10}$ n/cm$^2$s at the instrument PF1B \cite{Abele} of the ILL. The delays between activation and use of these sources were chosen sufficiently long to let parasitically activated isotopes (e.g. $^{38}$Cl, $^{69}$Zn) decay. 

\begin{figure}[h!]
	\centering
	\includegraphics[width=0.85\textwidth]{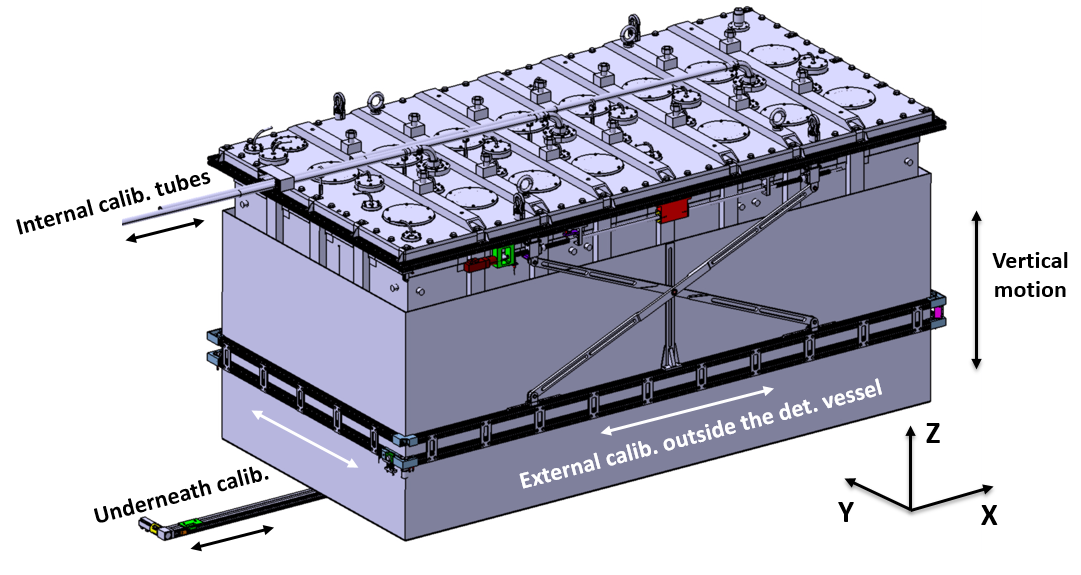}
	\caption{Calibration systems used in \STEREO{}.}
	\label{fig:calib_posit}
\end{figure}

 Radioactive sources can be deployed in or around the detector by three different methods, all with a positioning precision of better than 1 cm. Firstly, sources can be placed at any height in TG cells through vertical tubes. During phase-I, only the tubes of cells 1, 4 and 6 were accessible but starting from phase-II, tubes of cells 2 and 5 were connected. The tubes are centered in the transverse direction, but not in the long axis direction. They are 6 cm away from the adjacent cell in the reactor direction and 31 cm from the other adjacent cell. Secondly, a semi-automatized system enables the source positioning outside the detector's steel vessel at heights between 90 and 15 cm above the TG cell floor, and any position along the perimeter of the detector. Finally, another semi-automatized system underneath the detector vessel allows for source placement at any point below the cells along the central long axis of the detector. The three radioactive source calibration systems are sketched in Figure~\ref{fig:calib_posit}.

\begin{table}[h!]
	\centering
	\begin{tabular}{|c||c|c|c|c|c|c|c||c|}
		\hline
		Source  & $^{68}$Ge   & $^{124}$Sb & $^{137}$Cs & $^{54}$Mn  & $^{65}$Zn  & $^{60}$Co           & $^{24}$Na           & AmBe                 \\ 
		\hline
		{$\gamma$-ray energies} &  0.511 & 0.603    & 0.662 & 0.835 & 1.11 & 1.17  & 1.37  & 2.22 (H(n,$\gamma$)) \\
		 (MeV) 	& 0.511   &  1.69    &  - & -      & -   & 1.33 	  & 2.75    & 4.43 \\ 
		\hline
		Initial Activity (kBq)  & 90 & 2.4 & 37 & 90 & 3.3 & 50 & 5.9 & $250\cdot10^3$ ($^{241}$Am)  \\		
		\hline
	\end{tabular}
	\caption{$\gamma$-ray and neutron sources used for calibration of the \STEREO{} experiment. }
	\label{tab:set_sources}
\end{table}

\subsection{Electronics, DAQ and Monitoring}
\label{subsec:ElecDAQMonitoring}

A dedicated electronic system \cite{bourrion} was designed for the \STEREO{} experiment in order to trigger, process and read out the PMT signals.  
It is comprized of ten 8-channel front-end electronic boards (FE8), one trigger and readout board (TRB) and one board which drives the LED boxes used for the PMT calibrations. All these boards are contained in one single crate using the microTCA standard.
The first stage of the FE8 boards is an amplifier with two selectable gain factors: $\times 1$ for standard operation, $\times 20$ for single PE measurements.
PMT signals are then digitized at 250 mega samples per second, using a 14 bit ADC. 
The electronic noise, including the PMT bases and the FE8 analog part, is low enough to be sensitive to small PMT signals, from 10\,PE in gain $\times 1$ and 1\,PE in gain $\times 20$.
 After the ADC, a first-order high pass Infinite Impulse Response (IRR) filter ensures a null baseline and allows a faster baseline recovery after the detection of high energy events such as cosmic muons.

The FE8s are also in charge of a first level of trigger (T1). Samples are stored in a circular buffer to compensate for the latency associated with the trigger decision. An adjustable threshold is applied, with an OR condition, either to the signal amplitude (1 sample) or the charge (sum of several samples) of individual channels of the TG, GC and the muon veto. To avoid triggering on noise, the trigger can be applied to the numerical sum of 4 or 8 channels instead of individual channels. For TG and GC, PMT channels are clustered per cell. For the muon veto, channels of adjacent PMTs are clustered. 
During phase-I and phase-II, the T1 used for neutrino acquisition was applied on the charge, summed over 10 samples, of 4 summed clustered channels. For TG and GC, the T1 threshold was set to 1600~ADC units which roughly corresponds to a deposited energy in one cell of 300~keV. A threshold of 650~ADC units, corresponding to 32~PE, was also applied on the muon veto clusters to keep track of muons which only interact in the muon veto and not in the detector, in order to be able to reject delayed background events with offline analysis.
 
If T1 is accepted by the TRB, the event is processed. A constant fraction discriminator (CFD) algorithm is used to determine the pulse's start time on each individual channel, not only the triggered ones. Then, the total and the tail charge of each pulse are computed by summing the appropriate samples. The total charge is needed for the energy measurement while the tail charge provides pulse shape information for event discrimination.
In standard acquisition mode, only the processed data (CFD time, total and tail charges) for all channels are sent to the TRB in order to limit the data size and the dead-time due to data transfer. 
However, for debugging or pulse shape studies, samples can also be transferred. Typically, 60 samples (240~ns) are enough to fully contain pulses.
Thanks to the circular buffer, there is almost no dead-time associated with the processing. FE8 cannot trigger for a few clock cycles  (8 plus the number of samples transferred) after an event which allows for an instantaneous trigger rate of $3.6\cdot10^6$ s$^{-1}$.

From computed data, the TRB can apply a second level of trigger (T2) to reject unwanted events, such as those with too low or too high energy.  Any T2 condition can be implemented via the TRB FPGA.
During phase-I and phase-II, the T2 condition for neutrino acquisitions was rejecting events with too small signals in the three detector parts (lower than $\approx$200~keV in both TG and GC and lower than 70~PE in the muon veto). The later condition was set to reject $\gamma$-ray background coming from the neighboring instruments and generating small signals in the muon veto while keeping a good muon veto efficiency. The former condition was chosen lower than the T1 condition not to increase the energy threshold in the detector.
Finally, data of T2 events are sent to the computer disk. In addition to individual channel data, the event time is also registered with a 4~ns resolution. For this purpose, two clocks are used: the computer clock gives the time with a resolution of one second and the TRB 250~MHz clock gives a relative time between consecutive events with a resolution of 4~ns. Using a 32 bit counter allows to cover time differences of up to 17 seconds.
If the TRB FIFO is full, for example when the acquisition rate is higher than the data writing rate on the computer, T1 events are rejected. A counter allows measurement of the time during which the acquisition is frozen allowing for offline correction of dead-time.
For neutrino runs in standard acquisition mode, for which the acquisition rate varies from $1\cdot10^3$ to $2\cdot10^3$ s$^{-1}$ depending the background conditions, the resulting fraction of dead-time is negligible, less than 0.02\%.

All trigger and processing parameters (nature of T1, threshold value, number of samples for processing, T2 conditions,...) are adjustable and can be set via the data acquisition (DAQ) software. This software, which has been developed specifically for the \STEREO{} experiment within the NOMAD framework \cite{NOMAD} developed by the ILL, manages the acquisition and monitors the experiment. A variety of sensors have been deployed inside and outside the detector to continuously measure temperatures, absolute and relative pressures, liquid levels and magnetic fields.
PMT high voltages and currents, microTCA crate parameters and acquisition rates are also  monitored. All the values are stored in a database every minute. A homemade web software allows the vizualisation of all these observables.
Finally, two neutron detectors (BF$_3$ and  $^3$He proportional counters) are continuously running near the \STEREO{} detector to monitor the neutron background.  One is installed inside a 10~cm polyethylen shielding to be also sensitive to fast neutrons.

\section{Shielding}
\label{sec:shieldings}
Three principal external circumstances must be considered when designing shielding for a neutrino experiment at the \STEREO{} site:
(i) At the ILL, intense neutron beams are extracted from the reactor moderator to the experimental hall in order to perform neutron scattering experiments. This results in a high flux of thermal and fast neutrons (energy of about 25~meV and of several MeV, respectively)  and of neutron capture $\gamma$-rays (energy of up to 9~MeV).
(ii) The experimental site is not underground. Shielding against cosmic radiation is limited to about 15 meter water equivalent (m.w.e.) overburden at the \STEREO{} site by the transfer channel of the reactor (see Figure~\ref{fig:3DViewSite}).
(iii) Nearby neutron scattering experiments may employ magnetic fields of up to 15 T created by superconducting coils. The stray field at the \STEREO{} site can reach 1 mT.

\subsection{Passive radiation shielding}
\subsubsection{Background characterization}
\label{subsec:BackgroundCharacterization}

The radiation background conditions were measured in several campaigns in order to determine the appropriate shielding needed for \STEREO{}. Between these measurements, the shielding of neighbor instruments and beam casemates were reinforced based on the former measurements and the later campaigns served to validate and refine the shielding design.

Measurements of the neutron background were performed with a $^3$He tube (1 inch diameter, 36 cm long, 10 bars) placed in different configurations.
Measurements with the bare tube are sensitive to thermal neutrons. Fluxes of up to 100 neutrons/(cm$^2$ s) were observed, depending on the position and the configuration of neighboring instruments. The flux was the highest at the IN20 side of the \STEREO{} site. Note that this high thermal neutron flux complicates the $\gamma$-ray background measurements by $(n,\gamma)$ reactions close to the $\gamma$-ray detector.
Measurements with the $^3$He tube placed in the center of a polyethylene cuboid of $23 \times 23 \times 48$ cm$^3$ served to measure fast neutrons by moderating them in the polyethylene. The cuboid surface was covered by 4 layers of 5 mm boron-loaded rubber to block thermal neutrons. It could be used inside another outer cuboid of up to 4 layers of 5 cm thick polyethylene plates in order to vary the energy range of sensitivity to the fast neutron spectrum. Directional measurements were performed by removing all 4 polyethylene layers of one side of the outer cuboid. Spectrum information was also obtained using a NE213 scintillator detector. 
The measurements identified the following sources of fast neutrons (see also Figure~\ref{fig:2DViewSite}):
\begin{itemize}
  \item Collimation system of the GAMS instruments \cite{Koch1980} inside the beam tube H7: it contained apertures of $^6$LiF where neutrons of up to 16~MeV are produced by $(t, n)$ reactions \cite{Lone1980}. The collimation system was removed and replaced by an optimized beam plug, as explained in Section~\ref{subsec:ShieldingDescription};
  \item Casemate of the H13 beam port: the H13 beam port extracts an intense beam of thermal neutrons for the IN20 instrument, including a substantial fast neutron component. The casemate contains elements for beam definition, which scatter fast neutrons towards the \STEREO{} site;
  \item Reactor side of the \STEREO{} site: directional measurements after installing the beam plug still indicated an increased fast neutron source pointing towards the reactor, which could not be assigned unambiguously to sources in the H13 area or to the biological shielding of the reactor itself;
  \item IN20 instrumental area: fast neutrons were also detected along the wall between the \STEREO{} site and the IN20 area.
\end{itemize}

Gamma measurements were performed using a Germanium detector (Canberra 7229N or Ortec GEM 30210), with a 25\% efficiency relative to a 3 inch diameter and 3 inch long NaI scintillator, for a 1.3~MeV $\gamma$-ray energy. The efficiency of this Germanium detector for high-energy $\gamma$-rays, most relevant for the \STEREO{} background, is low. Therefore, this detector was mainly used to identify relevant materials for $(n,\gamma)$ reactions exploiting its high energy resolution (better than 0.3\% for the 1.3 MeV Ar transition) and for quantitative rate estimates. Positional and directional scans were performed using a NaI scintillator of 4 inch diameter and 6 inch length (Hawshaw 16S24/5A) with a poorer energy resolution (8\%{} FWHM at 1.3 MeV) but high efficiency for high-energy $\gamma$-rays. The measurements showed that the $\gamma$-rays partially originate from activation of surrounding materials by the ambient neutron flux and subsequent ${\beta}$-decay (e.g. ${^{40}{\rm Ar}(n,\gamma)^{41}{\rm Ar} \rightarrow ^{41}\rm{K}}$). However, at high energies, the $\gamma$-ray flux originates directly from neutron captures.
The main sources of high-energy $\gamma$-rays are:
\begin{itemize}
  \item $^{56}{\rm Fe}(n,\gamma)$ (7.631~MeV and 7.645~MeV) and  $^{27}{\rm Al}(n,\gamma)$ (7.724~MeV)
  $\gamma$-rays from neutron capture in structural material,
  \item decay of $^{16}$N ($\gamma$-rays of 6.1~MeV) in the primary cooling water circuit which passes below the floor of the \STEREO{} site. $^{16}$N is produced via $^{16}{\rm O}(n,p)$ when the heavy water is exposed to fast neutrons in the fuel element.
\end{itemize}

Both $\gamma$-ray and neutron measurements showed high reactor related background due to the close proximity of the reactor core and the presence of several neutron beam lines in the experimental hall.
Significant spatial and temporal variations in the $\gamma$-ray and neutron fluxes were also observed due to changes in experimental conditions of neighboring instruments.
To obtain an estimate of the true $\gamma$-ray flux present at the \STEREO{} site, an unfolding procedure has been developed based on the Van Cittert \cite{VanCit1931} and Gold's \cite{Gold1964} methods. This procedure corrects for the detection efficiency and the Compton scattering occurring inside the Germanium counter and provides the best estimate of the $\gamma$-ray flux for propagation through simulations of various shielding configurations. 
The \STEREO{} detector response was simulated using a GEANT4 code (see Section \ref{sec:MonteCarlo}). Assuming a signal of 400 neutrinos per day and a signal to accidental background ratio of 1.5, the maximum acceptable rates in the \STEREO{} detector were set to 350~s$^{-1}$ in the prompt energy window (2-10)~MeV and 1~s$^{-1}$ in the delayed energy window (5-10)~MeV.
This translates to large required attenuation factors for the shielding, from $2\cdot10{^{-5}}$ to $10{^{-6}}$ depending on the detector surface.

\subsubsection{Shielding description}
\label{subsec:ShieldingDescription}

The background measurements and rate estimates for the \STEREO{} detector resulted in a three-fold shielding concept: localized shielding applied to compact background sources, reinforcement of existing walls towards neighboring instruments and the dimensioning of the shielding in the \STEREO{} structure.

\begin{figure}[ht]
\centering
\includegraphics[width=0.65\textwidth]{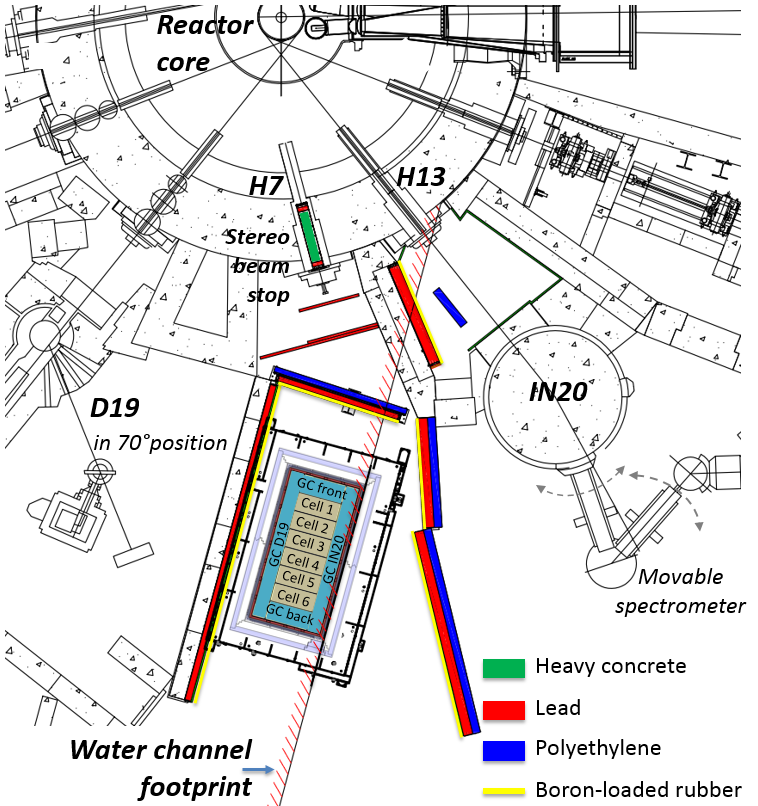}
\caption{Top view of the \STEREO{} site at level C of the ILL reactor. Reinforced shielding along the side walls and in the front area is highlighted.}
\label{fig:2DViewSite}
\end{figure}

As the \STEREO{} area is located at the exit of the H7 tube that crosses the reactor pool (see Figure~\ref{fig:2DViewSite}), a series of MCNPX simulations were performed to design an adapted plug for the tube. In the final design, the plug consists of, as seen from the direction of the reactor, 2~cm of Boral, 8~cm of lead, 1~m of heavy concrete and 8~cm of lead. The H13 casemate bordering on \STEREO{} was reinforced by borated polyethylene close to the neutron beam to protect against scattering from the beam-defining elements. The thickness of an existing lead wall inside the H13 casemate toward \STEREO{} was increased from 10 to 20~cm and the casemate was covered with boron-loaded rubber from the inside.

To protect the \STEREO{} area from the reactor and neighboring instruments, additional delimiting walls were installed  (see Figure~\ref{fig:2DViewSite}). The front wall between the \STEREO{} area and the reactor consists of 10 cm of polyethylene and 10 cm of lead. On the D19 side, 10~cm of a 40~cm concrete wall were replaced by 10~cm of lead whereas on the IN20 side a 15~cm lead wall was doubled with 15~cm borated polyethylene. All walls are 2.4~m high, shielding the detector volume but leaving the muon veto (placed at 3~m height) unprotected. The mechanical frames of the walls were covered by boron-loaded rubber to reduce neutron capture in the steel.

\begin{figure}[h!]
\centering
\includegraphics[width=0.60\linewidth]{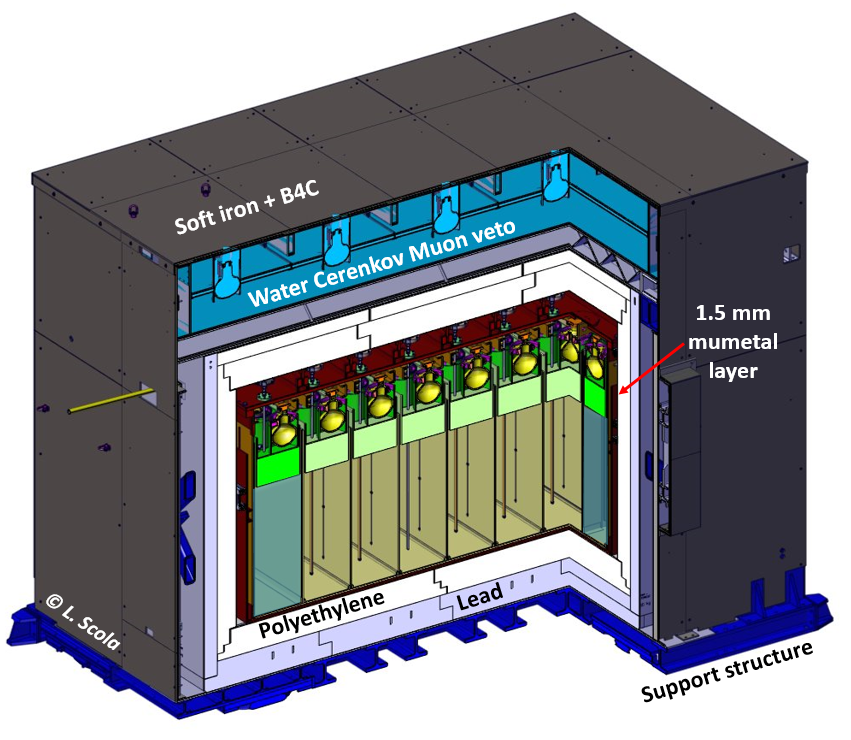}
\caption{Cut view of the \STEREO{} assembly held by the support structure illustrating the layers of the inner shielding (see text for details).}
\label{fig:Det_Assembly_CutView}
\end{figure}

A passive shielding of about 65~tons made of 5\% borated polyethylene and a lead outer layer encloses the neutrino detector to protect it from the external background and from cosmic muon induced background (see Figure~\ref{fig:Det_Assembly_CutView}). Borated polyethylene is used to moderate and capture neutrons while the lead serves to mitigate the $\gamma$-ray background in the hall. 
The \STEREO{} site can accommodate such a heavy structure since the maximum floor load of the area is 10~t/m${^{2}}$, the largest in the reactor building. 
The polyethylene shielding (total 6.1~tons) is thicker above and underneath the detector than on the side walls. 
The layer on the top is 297~mm thick in order to protect the inner detector from fast neutrons produced by cosmic muons in the top lead layer and in the transfer channel, while the layer thickness is 200~mm underneath the detector.
On the side walls, the thickness of polyethylene is only 147~mm as the outer crown of the Gamma-catcher serves as an additional shielding layer against neutrons. 
With at least 10~cm thickness in each direction, the outer lead layer (total 58.9~t) serves to mitigate the residual $\gamma$-ray background in the \STEREO{} area. 
As high energy $\gamma$-rays from the decay of $^{16}$N passing through the floor of \STEREO{} have been observed, the thickness of the lead layer below the detector is 20~cm. 
Similarly, the thickness of the top layer is 15~cm as there were indications that neutron captures occur in the heavy concrete of the transfer channel. 
The lead is supported on the outside by a strong structure, based on a girder frame anchored to the floor, which guarantees the mechanical stability of the shielding and fulfills the requirements of resistance to earthquake. 
Three apertures were realized in the hermetic shielding: two small ones serve to pass through the shielding cables, inner calibration tubes and the rail of the underneath calibration system; the third one is a trapezoidal aperture, closed by a sliding door with the same shielding as the walls, allowing access to the outer calibration system.
 The 25.6~cm deep veto water tank (see Section \ref{subsec:veto}) serves as an additional absorber above the shielding.

\subsubsection{Single rates}
Table~\ref{tab:GammaRates} compares the rates estimated from background measurements and simulations as explained in Section \ref{subsec:BackgroundCharacterization}, with the actual rates measured in the \STEREO{} detector and the rates which were considered for the experiment proposal before the measurements on site. The uncertainties of the estimated rates are dominated by the extrapolation of scans with a small Ge detector to the large volume of the \STEREO{} detector, the propagation through the massive \STEREO{} inner shielding and the dependence of the $\gamma$-ray background on the configuration of the surrounding instruments. Furthermore, neutrons passing through the thermal neutron shielding may create $\gamma$-rays inside the \STEREO{} inner shielding. 

Finally, the single rates measured in the prompt energy window are much lower than the rates considered for the experiment proposal while the rates in the delayed energy window are comparable. The product of these two rates drives the rate of accidental coincidences in the detector. It stays well below the specification of the proposal confirming that the accidental background will not limit the sensitivity of the experiment.

\begin{table}[h!]
\begin{center}
\begin{tabular} { l | c |  c| }
\cline{2-3}
   & Prompt Energy window  & Delayed energy window  \\
   & (2 - 10) MeV & (5 - 10) MeV \\
\hhline{~==}
\hhline{-~~}
\multicolumn{1}{ |c|| } {Estimated rates (s$^{-1}$)}& 25  & 0.4  \\
\multicolumn{1}{ |c|| } {Measured rates (s$^{-1}$)}  & 13 & 3.5  \\
\multicolumn{1}{ |c|| } {Proposal rates  (s$^{-1}$)}  & 350 & 1 \\
\hline
\end{tabular}
\caption{Comparison between expected and measured rates in the \STEREO{} experiment for the two energy windows of interest. Measured rates are without muon related events. These rates can vary with the experimental conditions of neighboring instruments, maximum values are reported.}
\label{tab:GammaRates}
\end{center}
\end{table}

\subsection{Muon veto}
\label{subsec:veto}
\begin{figure}[h!]
	\centering
	\includegraphics[width=0.5\textwidth]{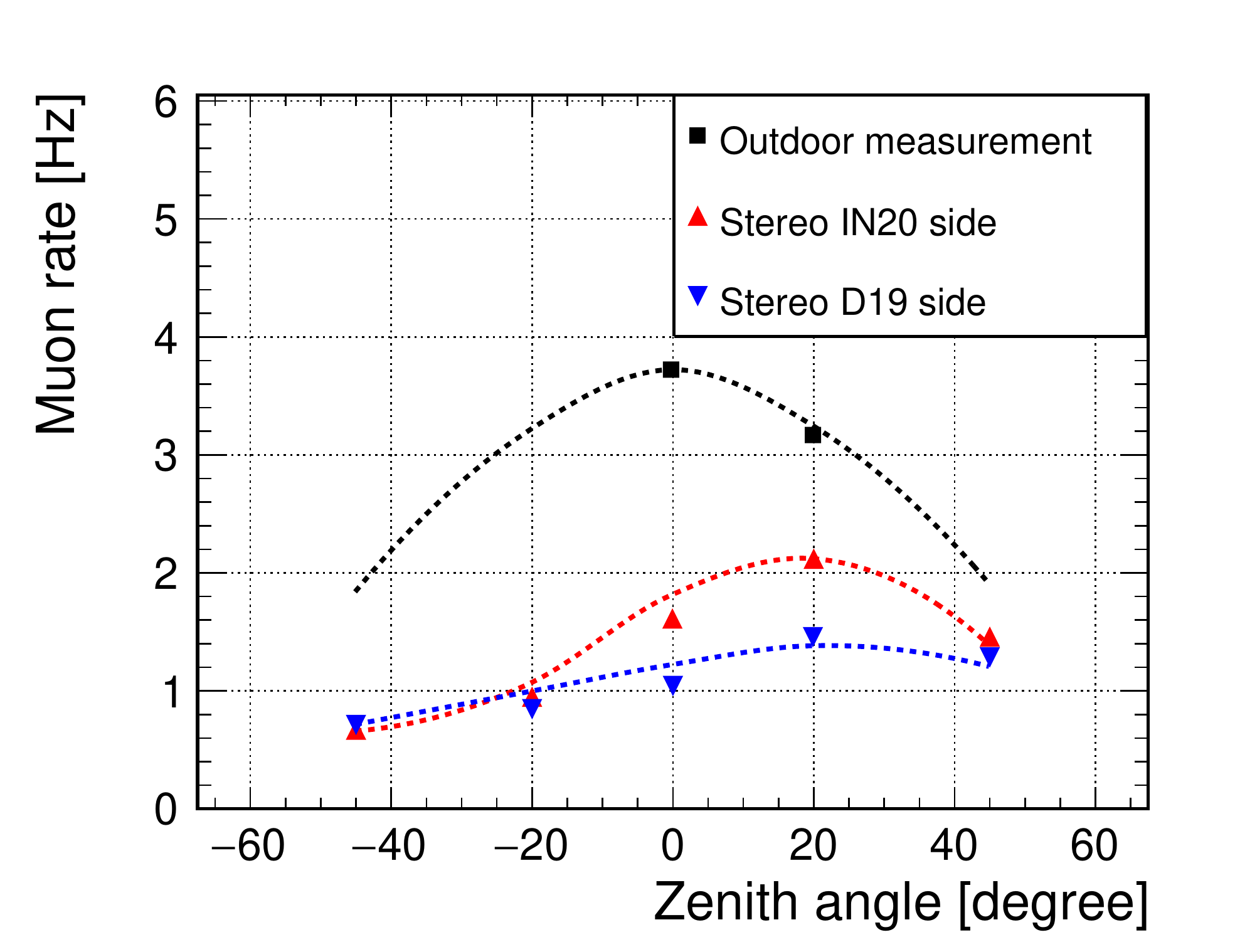}
	\caption{Muon fluxes at the ILL for different locations and zenith angles. Markers correspond to measurements, dashed lines are the simulation results. The simulation flux normalization is done once using the outdoor measurement at 0~degree zenith angle. }
	\label{fig:MuonFlux}
\end{figure}

Although the reactor building and the transfer channel above the detector already provide a shielding of 15 m.w.e., the cosmic radiation still constitutes an important source of background. 
The muon flux for varying incident angles was measured at different positions on the \STEREO{} site and outside the building using a coincidence between 2 plastic scintillators (52~cm long, 12~cm wide and 0.8~cm thick). They are stacked above one another with a 20~cm distance. This allowed quantification of a shielding factor for vertical muons up to 4, from which the value of 15 m.w.e was inferred. As \STEREO{} is located at one side of the transfer channel, the shielding factor depends on the position on the \STEREO{} site and on the incident angle as presented in Figure~\ref{fig:MuonFlux}. The cosmic rates have been simulated using the CRY cosmic generator \cite{cry17} and a complete description of the ILL building. Since the CRY generator provides tabulated rates at sea level, a common normalization factor has been determined using the outdoor scintillator measurement and applied to all simulated positions.

The neutrons and $\gamma$-rays produced by the muon spallation can mimic the IBD signature by forming correlated signal coincidences. One of the countermeasures implemented against this background is the use of a water Cherenkov muon veto on top of the \STEREO{} shielding. The dimensions of the active part ($396 \times 240$ cm) were chosen to geometrically cover the detector and its shielding. The muon veto tank is filled with 25.6~cm of demineralized water and a wavelength shifter (4-Methylumbelliferone (4MU)) to absorb the Cherenkov photons in the UV range and re-emit them around 420 nm, where the PMT quantum efficiency reaches a maximum of $\simeq 25\%$.
Several concentrations of 4MU have been tested from 2 to 10~PPM with a linear increase of the detected optical photons at low concentration and a saturation for higher values. A concentration of 6~PPM was chosen and tested over more than a year in a veto prototype with no visible aging effects on the detector response. Optical photons are reflected on all sides of the muon veto tank by a highly diffusive sheet of Tyvek 1059B in order to increase the total collected light on all the PMTs and to smooth the detection efficiency over position.  
The light collection is performed by 20 Hamamatsu photomultipliers (R5912) with the photocathode immersed in the water. The PMTs' gains are monitored and adjusted using the same LED calibration principle as in the \STEREO{} detector. A hard trigger on the charge summed over a bloc of 4~PMTs (see Section~\ref{subsec:ElecDAQMonitoring}) has been preferred to a single PMT trigger as it reduces the $\gamma$-ray contamination. During normal operation, the average number of PEs reaches 530 for vertical muons.

After correcting for the effect of atmospheric pressure on the cosmic ray flux, the muon rate in the veto reaches an average of 660~s$^{-1}$ at 1020 hPa with some time variation of a few percent correlated to water level changes in the reactor core pool and the transfer channel.The muon veto efficiency is monitored using neutrino runs which contain a large amount of cosmic particle events. The veto efficiency is defined as the probability to detect a muon that traverses the veto. Muons passing through the veto are identified by selecting vertical muons in the detector, i.e. muons that only deposit energy in one cell.
Since a muon deposits around 2~MeV/cm in the LS at its ionization minimum, it only needs a few centimeters of path in the detector before it induces the saturation of all the PMTs in one cell, providing an easy identification and selection of vertical events.
A follow-up study since data taking started showed a veto efficiency of $(99.5 \pm 0.1)\%$. The difference in veto efficiency between the reactor-on and reactor-off periods is lower than 0.1\%, the corresponding systematic uncertainty on the neutrino rate is thus negligible.

\subsection{Magnetic shielding}
\label{subsec:MagneticScreen}

The magnetic field generated on the \STEREO{} site by superconducting magnets of the IN20 instrument could only be measured after finishing the design phase of \STEREO{}. The magnetic shielding design was therefore based on simulations using the two complementary codes Comsol \cite{COMSOL} and MAENTOUCH \cite{Maentouch}. The latter code is a homemade program for solving magnetostatic problems using the boundary element method. It can handle 3D magnetic field calculations involving field sources and possibly ferromagnetic materials.
Without \STEREO{}'s magnetic shielding, the measured maximum stray field from the IN20 magnet operated at 13.5 T was 0.41 mT (0.42 mT) in the detector (veto) PMT plane, 18\%{} (30\%{}) smaller than expected from the simulations. The field decreased more quickly with distance to IN20 than in the simulations. Both effects may be explained by reinforced concrete around the \STEREO{} site that was not included in the simulations.

The magnetic shielding designed for the \STEREO{} experiment consists of several layers of different materials. First, walls of soft iron (10~mm thick) were installed in a geometry that completely surrounds the \STEREO{} shielding structure. 
A layer of mu-metal (1.5~ mm) was installed forming again a closed volume around the detector vessel. 
For the deployed Hamamatsu PMTs, 2 concentric cylinders of mu-metal (1 mm thick each) were placed around the photocathode region of the Veto PMTs, and a single cyclinder was placed around the TG/GC PMTs.
Magnetic shielding of the large photocathode PMTs is essential to minimize the loss of PE along their trajectory to the first dynode.

The magnetic shielding of \STEREO{} was demonstrated to be effective and to fulfill the requested requirements during periods when the IN20 spectrometer was operated at high magnetic fields. For the muon veto, the PMT signal amplitudes varied up to 20\% for the most exposed PMTs along the IN20 instrument. However, their variations did not induce any efficiency loss thanks to the low thresholds allowed by the use of a second level of trigger. For the TG and GC, the requirement is more stringent as a magnetic field change at the PMTs would directly influence the energy measurement. As expected, no effect could be seen on the PMT signals (within the precision of the monitoring estimated to be much better than 1\%) even for the PMTs close to the IN20 instrument. 

\section{Monte Carlo Simulation}
\label{sec:MonteCarlo}

Simulations of \STEREO{} were performed using the Geant4 toolkit (version 10) \cite{geant4}. 
A complete description of the detector, its shielding, the calibration systems and the muon veto was implemented. It includes all the main mechanical parts, liquid scintillators and photomultipliers. All physical processes are simulated from the particle interactions to the light collection to the PMTs. A particular attention was paid to the description of the scintillation liquid properties and of the reflective walls in the TG and GC cells. A simple and approximative optical model was implemented in the simulation to reproduce the effects of the loss of reflectivity due to the LS leakage into the reflective walls. The model and its parameters were adjusted for the different cells by comparison with experimental data.
The scintillation light emission parameters, such as the light yield, were also tuned as described in Section~\ref{sec:Erec}.
The final step in the simulation reproduces the PMT signals including the electronics response. The simulation software can produce output data in the same format as real data in order to be analysed with the same codes.


\section{Detector Response}

\subsection{PMT calibration}
\label{subsec:PMTCalibration}
\begin{figure}[h!]
	\centering
	\includegraphics[width=0.65\textwidth]{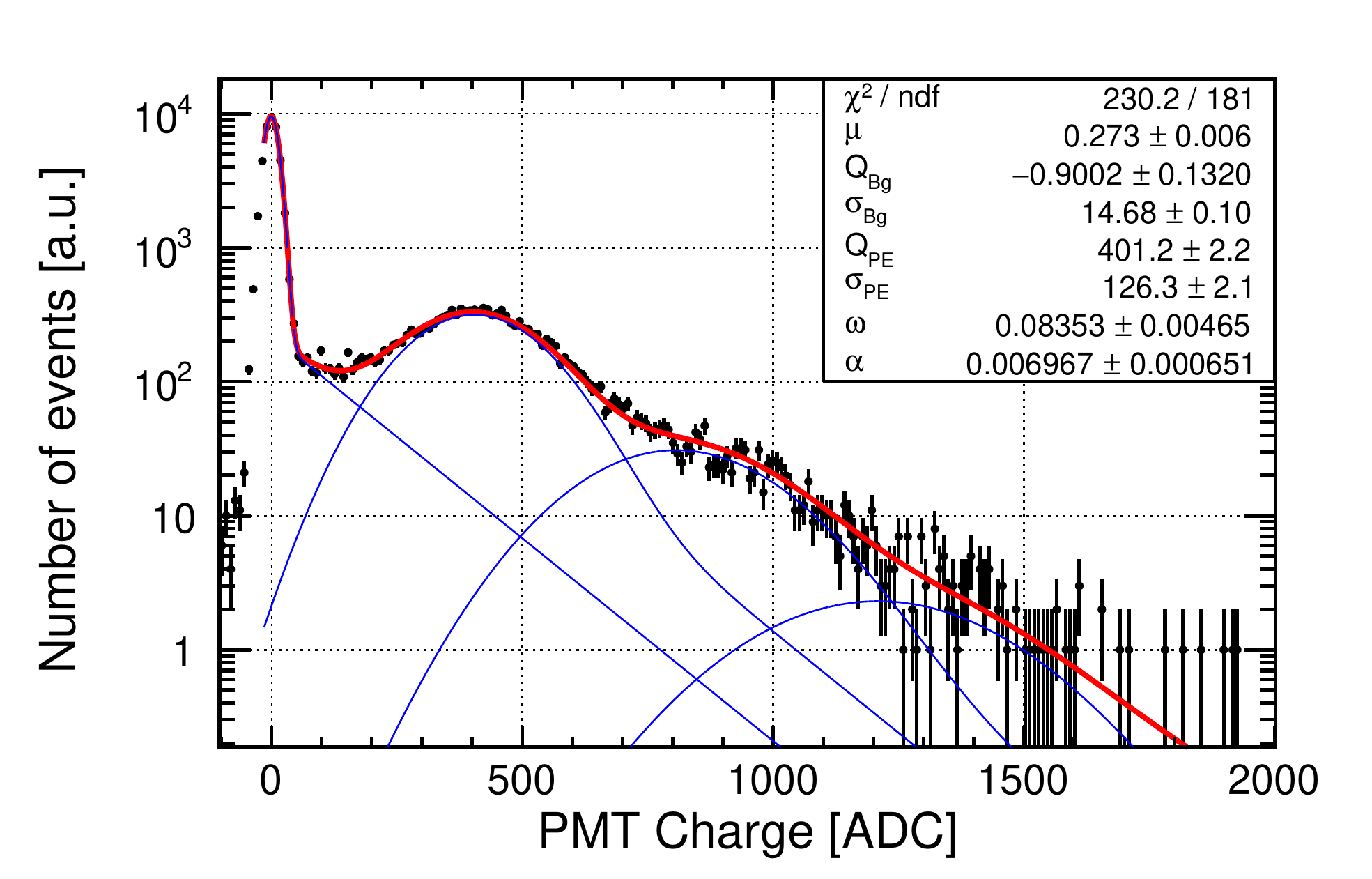}
	\caption{Typical single PE spectrum from a Hamamatsu R5912-100 PMT in a TG cell fitted with a sum of functions which parametrize the noise and the PMT response. }
	\label{fig:SPEspectrum}
\end{figure}

Relative PMT gains are measured from single PE spectra obtained by illuminating the diffusor balls with the LED light sources operated at low intensity as explained in Section \ref{subsec:calibration}. 
The charge integration window, triggered by the synchronization signal of the LED generator, is set at 20 samples (80~ns)  (see Section \ref{subsec:ElecDAQMonitoring}) to optimize the signal over background ratio. The charge is recorded using the gain $\times$20 setting of the FE board.
Measurements are carried out every 2 hours during short periods of 30~s.
An example of charge histogram is shown in Figure~\ref{fig:SPEspectrum}. 

The PE peak is clearly visible on the right of the pedestal peak and multiple PE peaks are also visible at higher charges.
The measured single PE spectrum is modeled \cite{bellamy} by the convolution of the response functions of the PMT with background processes. PMT response is treated as two independent parts: the photo-conversion and electron collection at the first dynode, and the amplification of the initial charge emitted by the photocathode. Background processes are also split into two groups: the pedestal which is present even in the absence of light, and small amplitude signals induced by spurious phenomena in the PMTs following the photon detection (such as the photoelectric effect on the first dynode or thermionic emission).  The fit parameters describing both the background and the signal are given in Figure~\ref{fig:SPEspectrum}.
The fitting procedure finds very good agreement between the data and the model. The position of the single PE peak, ${Q_{PE}}$, is defined with an accuracy of the order of 0.5\%. 
The resolution of the single PE ${\sigma_{PE}/Q_{PE}}$ is of the order of 30\% and the peak to valley ratio is about 3. 
  
\subsection{PMT and electronics linearity}
\label{subsec:Linearity}
The PMT and electronics linearities are measured using 4 LEDs. During dedicated runs, LEDs are switched on periodically with frequencies which are multiples of each other allowing to obtain all possible combinations. PMT signals with several LEDs simultaneously switched on are compared to the sum of PMT signals with only one LED switched on, which serves as a reference. In the case of a perfect linearity, the two terms are equal.
This relative method allows a better precision than an absolute calibration using only one LED at various amplitudes.
The whole range of interest up to 1500~PE per PMT can be covered by repeating the measurement and varying the LED amplitudes. A specific analysis method has been developed to take into account the fact that PMT signals with only one LED switched on can be already affected by a non-linearity. In order to allow the convergence of the method, PMTs and electronics are assumed to be perfectly linear below 200~PE. 

\begin{figure}[h!]
	\centering
	\includegraphics[width=0.65\textwidth]{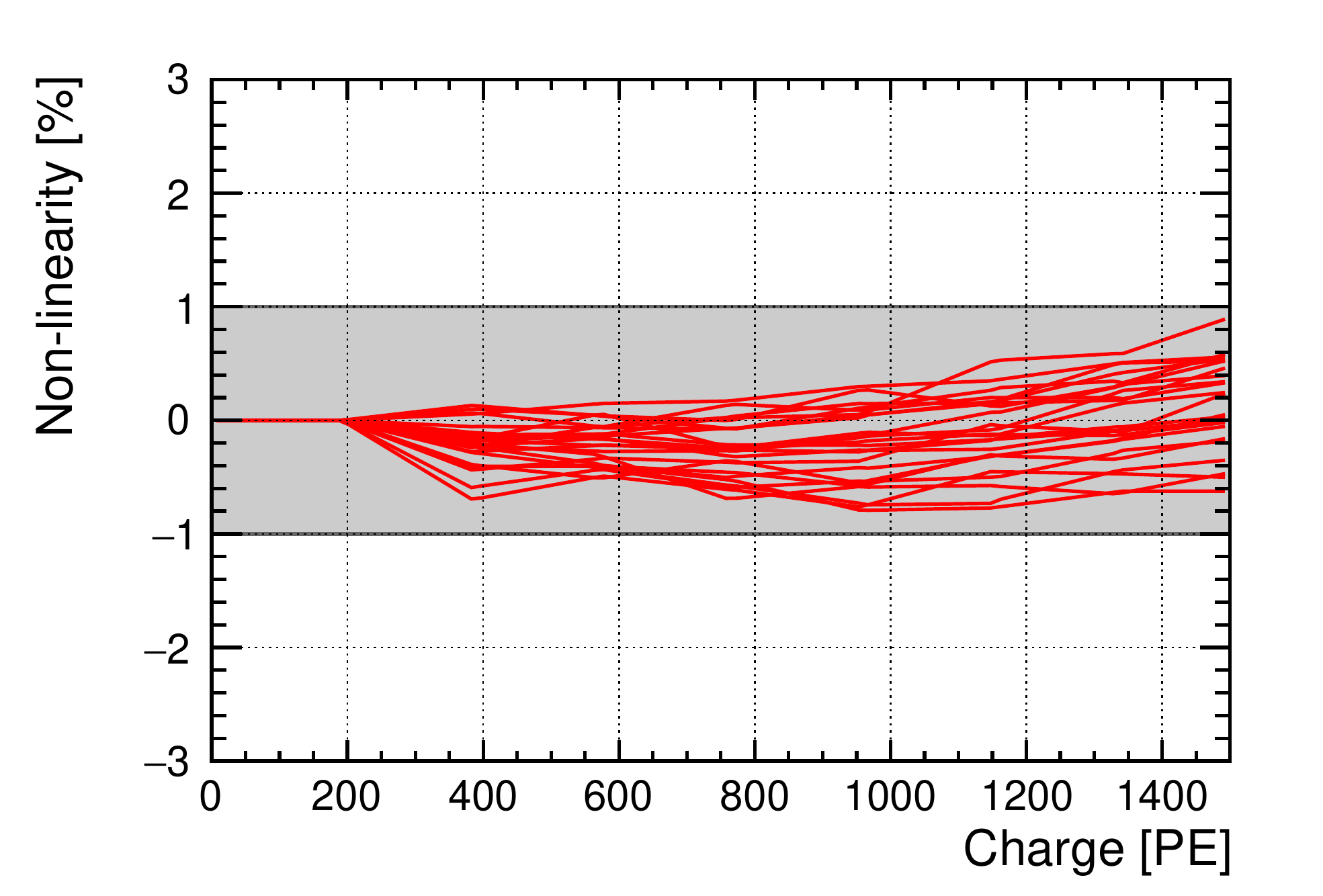}
	\caption{Deviation from linearity for all TG PMTs. PMTs and electronics are assumed to be perfectly linear below 200~PE.}
	\label{fig:PMTLinearity}
\end{figure}

For all the detector PMTs, the deviations from linearity are lower than 1\%, as shown on Figure~\ref{fig:PMTLinearity} for the TG PMTs fulfilling the experimental requirements.

\subsection{Light cross-talk between cells}
\label{subsec:LightCrossTalk}
Keeping track of the light cross-talk between cells is crucial for the \STEREO{} experiment in order to assess the response of the detector.
The light cross-talk coefficients from cell $i$ to cell $j$, $ L_{ij}$, where $i$ and $j$ correspond to every TG and GC cell, are defined as the ratio of collected charges in cells $j$ and $i$ for single-cell energy deposits in cell $i$:
\begin{equation}
  \label{eq:crosstalk_coefficient}
    L_{ij}=\frac{Q_j}{Q_i}
\end{equation}

\begin{figure}[h!]
	\centering
	\centering
	\subfigure[]{\includegraphics[width=0.48\textwidth]{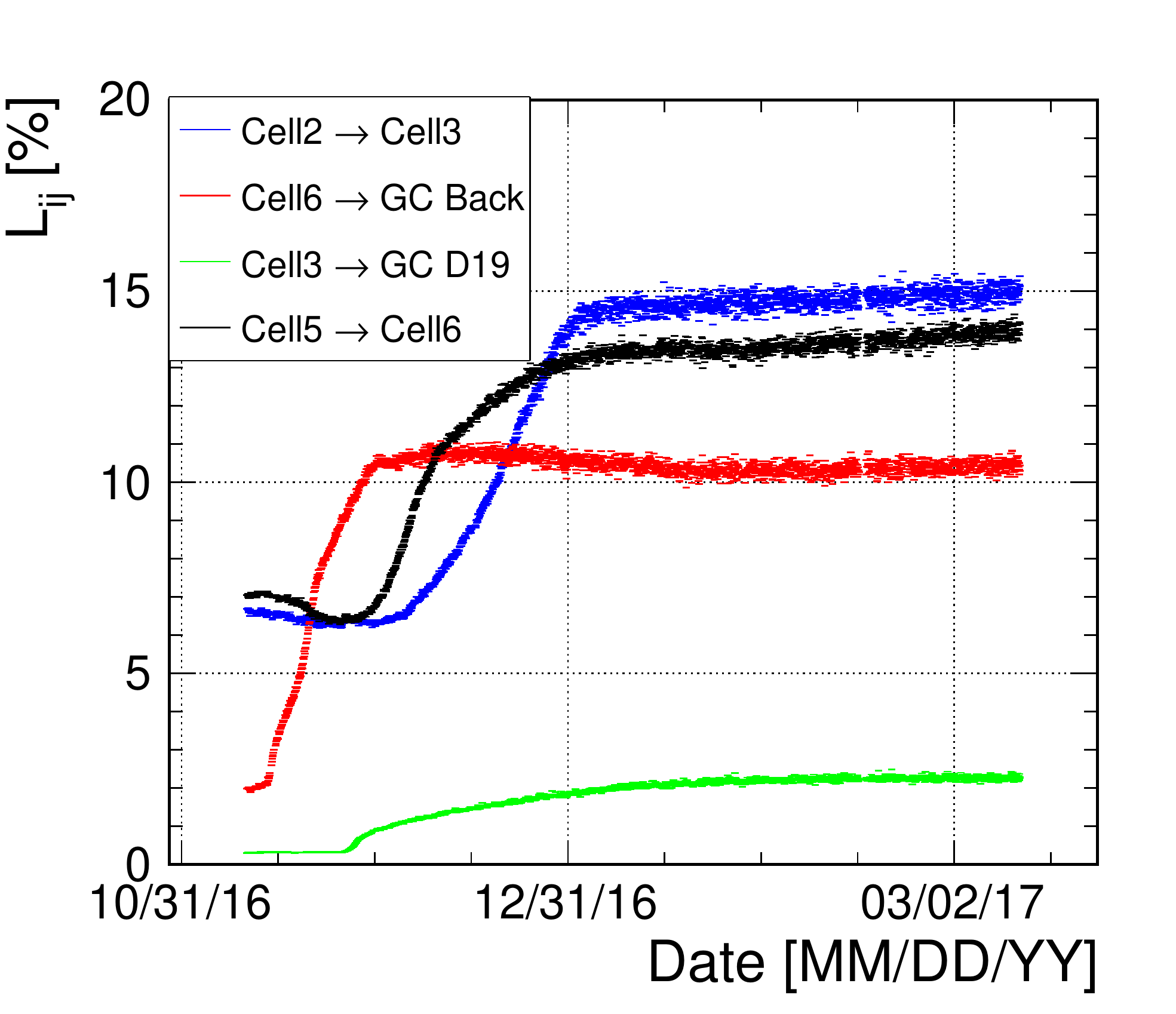}}\quad
	\subfigure[]{\includegraphics[width=0.48\textwidth]{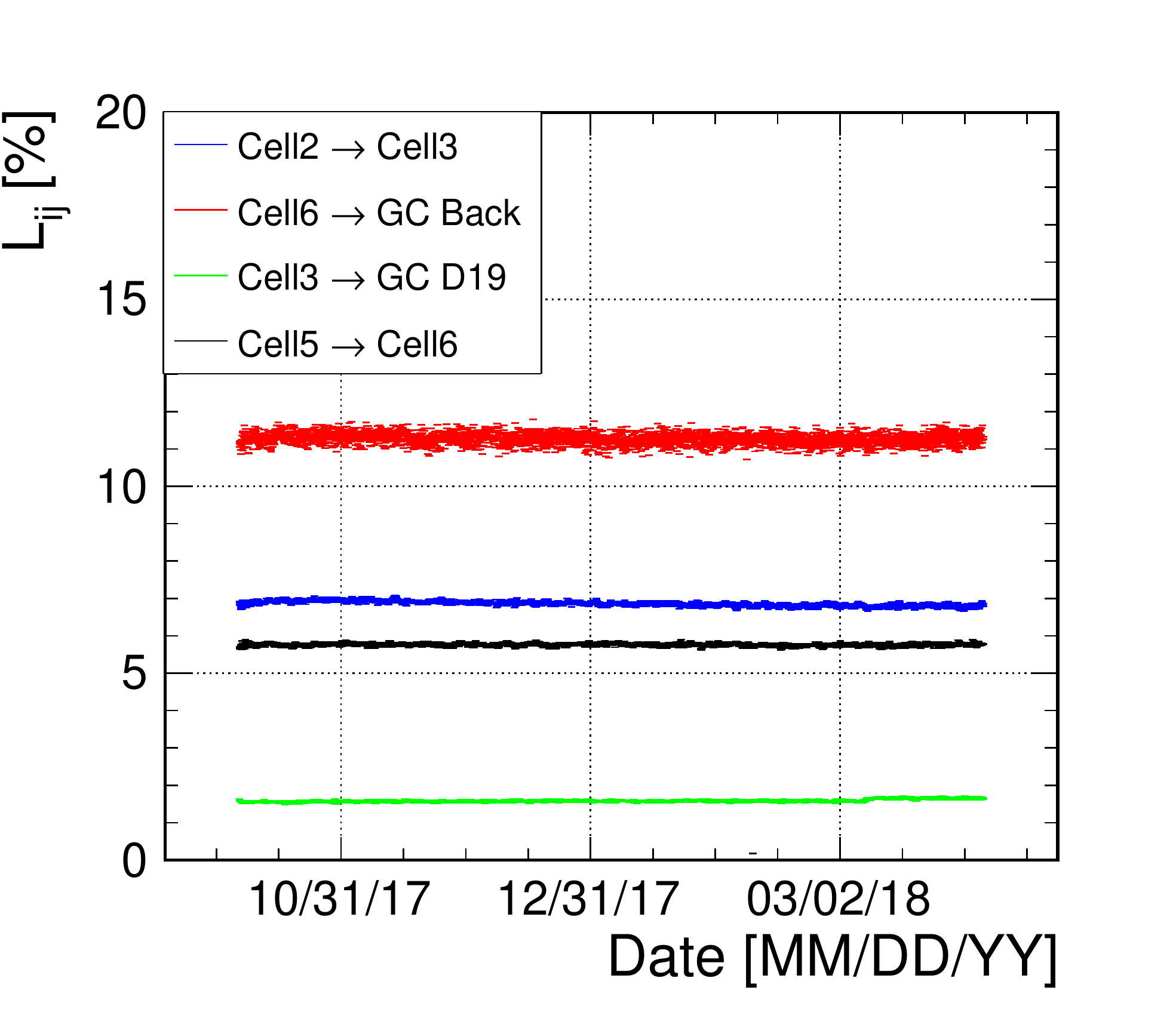}}\\
	
	\caption{Examples of light cross-talk evolution during the phase-I (a) and the phase-II (b) of the experiment for selected $L_{ij}$ coefficients. The repair of reflective walls separating the TG cells reduced the light cross-talks between TG cells in phase-II. }
	\label{fig:LLEvolution}
\end{figure}

By definition, all $L_{ii}$ are equal to 1. Note that this definition does not yield the physical light cross-talk as $Q_j$ depends on the light collection efficiencies of the cells.
Two independent methods, yielding consistent results, have been developed to measure the $ L_{ij}$ matrix elements by selecting single-cell energy deposits. A first method is based on full energy deposition events from calibration runs performed with a $\gamma$-ray source, namely $^{54}$Mn, while the second one uses high energy events ($\simeq$ 10 to 40 MeV) induced by cosmic-rays recorded during standard runs.
The first method has the advantage of being sensitive to the vertical variation of the cross-talk by using different source positions along the $z$-axis. The mean  light cross-talk of a cell is then obtained by averaging all values. 
The second method uses events which are more homogeneously distributed within each cell. It can also be applied on neutrino runs, which allows to follow the evolution of light cross-talks continuously, on a hourly basis, as illustrated in Figure~\ref{fig:LLEvolution}, without the need of specific calibration runs. 

During phase-I of the experiment, the unexpected leakage of LS inside the air gap of the reflective walls induced an increase of light cross-talk between cells (see Section \ref{subsec:DetectorDesign}) visible by the steps in  Figure~\ref{fig:LLEvolution}(a).
This effect prompted the development of a non-trivial method to calculate the energy deposited by a particle in the detector, as will be discussed in Section~\ref{sec:Erec}.
The light cross-talks were in general lower and much more stable during phase-II, after the repair of the reflective walls separating the TG cells.

\subsection{Cell-by-cell calibration}
\label{subsec:cell-by-cell-calib}

The mono-energetic $\gamma$-ray source $^{54}$Mn has been extensively used for the energy scale calibration of the \STEREO{} detector and to check its response stability. By selecting full energy deposition (FED) events in a given cell $i$, the sum of the PMT charges can be simply translated from PE to MeV units by means of calibration coefficients $C_i$. 
The $C_i$'s of the TG cells have been derived by deploying the source through the internal calibration tubes. To evaluate the dependence on the distance between the energy deposit and the PMTs, the source was placed at five different heights for each of the cells with a calibration tube. The GC cells can be calibrated with sources either on the outside (external calibration system) or in the internal calibration tubes, thanks to the activity of the $^{54}$Mn source which is high enough not to be limited by the background and to the interaction length of the 0.835 MeV $\gamma$-rays (10 cm in LS).

By definition, $C_i$ is calculated as $C_i$ = $Q^{\text{FED}}_{i}$/$E_{\text{dep},i}$ where $Q^{\text{FED}}_{i}$ is the mean total charge of the cell \textit{i} PMTs for FED events in the cell $i$ and $E_{\text{dep},i}$ is the expected mean deposited energy in the cell $i$ obtained from Monte Carlo (MC) simulations. The energy $E_{\text{dep},i}$ is a few percent smaller than the nominal $\gamma$-ray energy due to energy losses by Compton scattering in detector dead zones, such as the source holder, the calibration tube and the acrylic walls of the cell. For TG cells, the difference is less than $4\%$. For GC cells, the difference is $\sim 9\%$ for central positions and $\sim 14\%$ for corner positions. These larger deviations are due to the larger amount of inert material that $\gamma$-rays can meet in comparison with TG cells.

\begin{figure}[ht]
\centering
\includegraphics[width=0.65\textwidth]{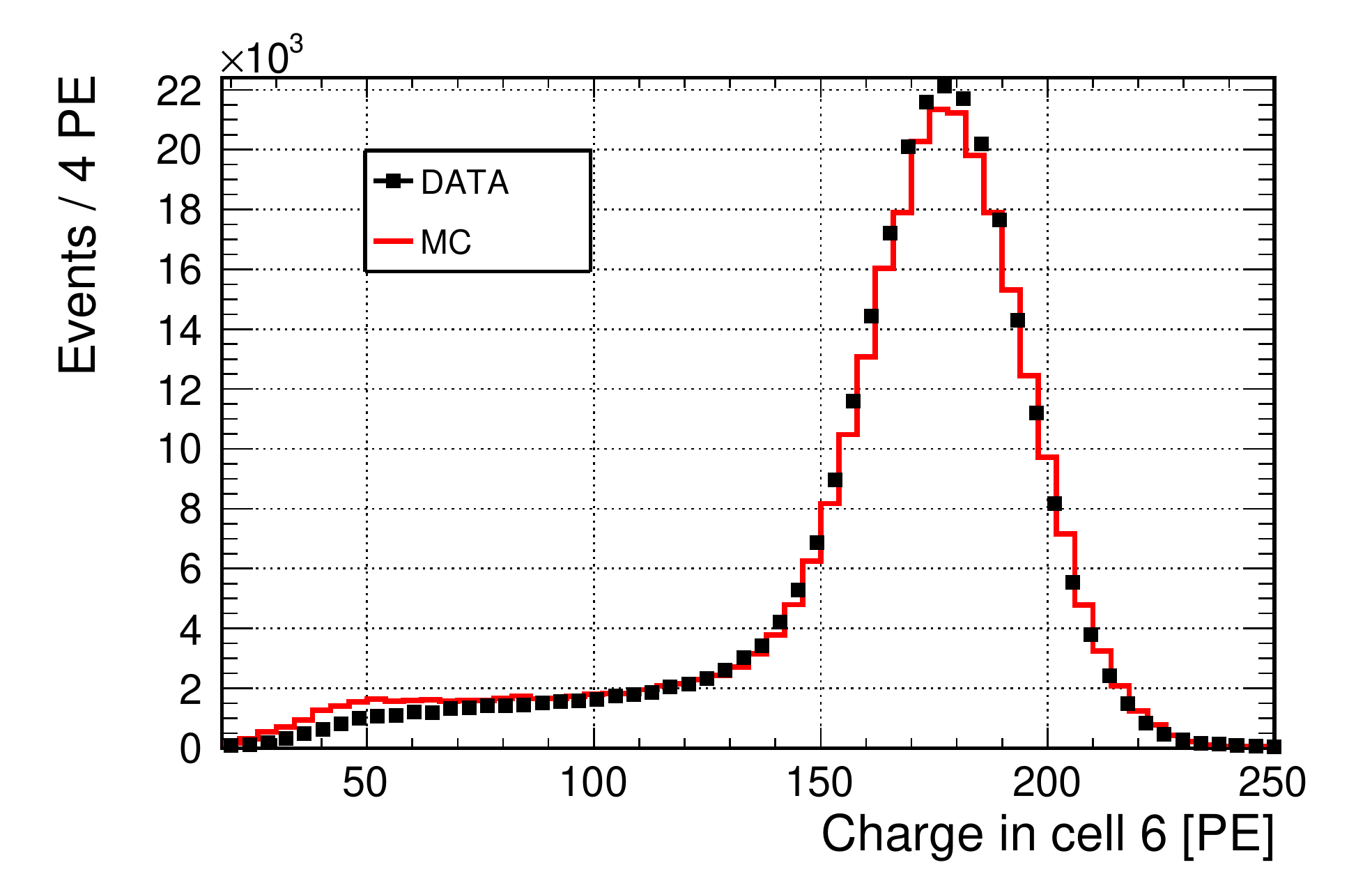}
\caption{Comparison between data (black) and Monte Carlo (red) charge spectra in cell 6 when the ${}^{54}$Mn source is deployed at half of the cell height. Similar plots are obtained for all the cells.}
\label{fig:MnSpectra}
\end{figure}

Because of the light cross-talk between cells, the neighboring cells observe some light signals even in the case of FED events.
To take into account this effect in the FED selection, the measured charges $Q_j$ in the four closest cells to the cell $i$ are required to be lower than $Q_i \cdot c \cdot L_{ij}$ where $c$ is a tuned parameter (a typical value is 2) which controls the tightness of the cut. 
This condition is applied both on data and MC simulations.
Figure~\ref{fig:MnSpectra} compares the data and MC charge spectra with such a selection, when the source is placed at half of the cell height ($z = 45\,$cm). The agreement is very good, especially for the full energy peak. 

The mean values of $Q^{\text{FED}}_{i}$ and $E_{\text{dep},i}$ can be obtained from data and MC energy spectra by fitting the full energy peak. Independent analyses using slightly different selection conditions and fitting ranges yield consistent $C_i$ coefficients demonstrating the robustness of the method. Thus, coefficients at 0.835~MeV vary from 220 to 260~PE/MeV for the TG cells, except for the cell 4 where the coefficient during phase-I was 2.5 times smaller due to the loss of the optical coupling (see Section \ref{subsec:DetectorDesign}).
Concerning GC volume, long cells show lower coefficients (270 to 360 PE/MeV) than the short ones (330 to 400 PE/MeV) as expected from their lower coverage of photocathode per unit surface. As for the TG cell 4, the coefficient of the GC front cell was smaller during phase-I, by a factor of 2.2.

\begin{figure}[ht]
\centering
\includegraphics[width=0.65\textwidth]{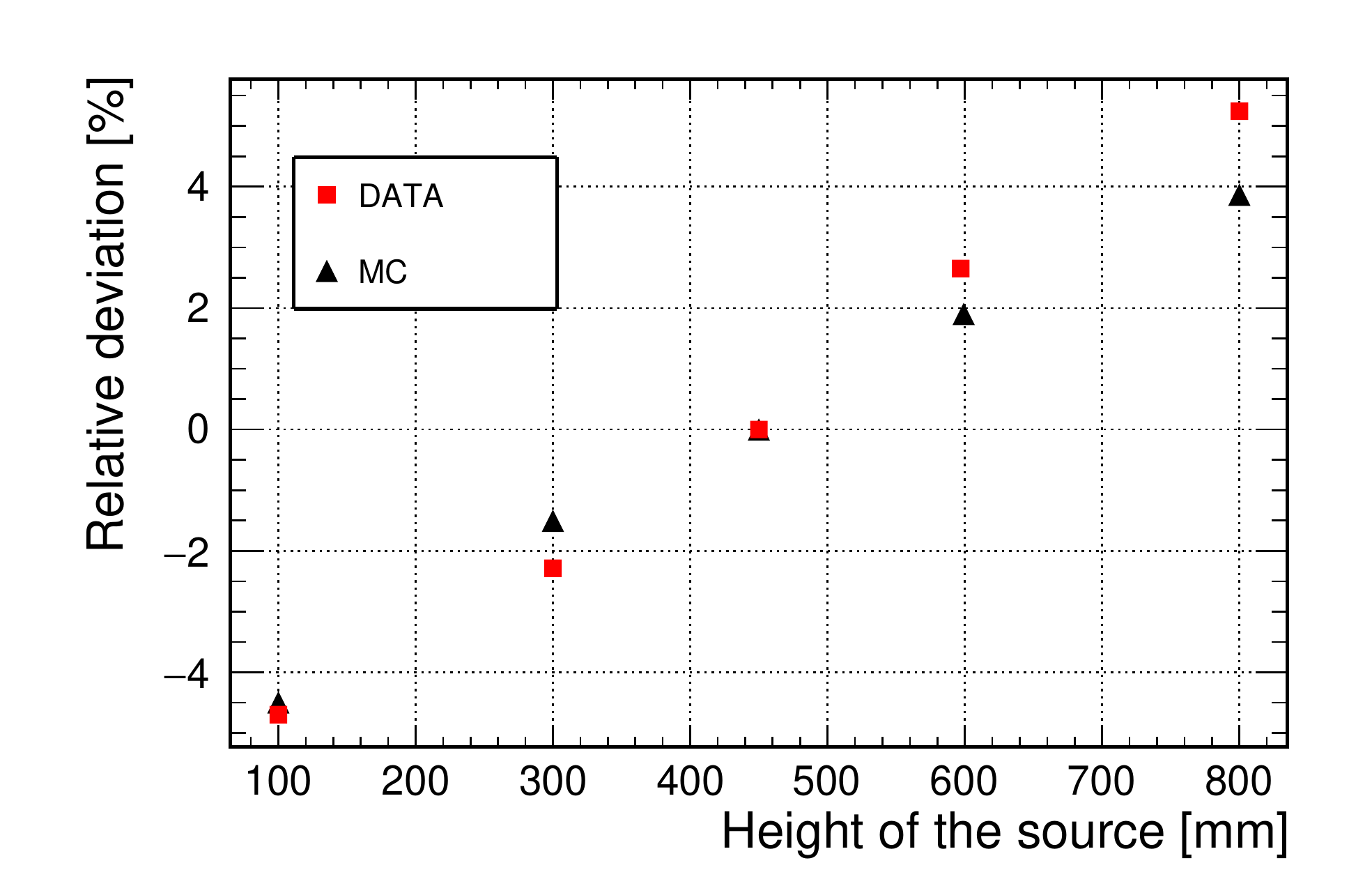}
\caption{Deviation of the charge peak position at different deployment heights relative to the charge peak at $z = 45\,$cm, for TG cell 6.}
\label{fig:ZDependence}
\end{figure}

Figure~\ref{fig:ZDependence} illustrates the deviation of the charge peak position for a source deployment at different heights relative to the cell charge peak at the half of the cell height. Data and MC show a good homogeneity of the energy response throughout a cell as the response varies by only $\pm\,4\,$\% between the top-bottom edges and the center of the cell. Data and MC agree at the sub-percent level across most of the vertical range, percent level discrepancy is only observed for the top 10~cm of the cell.  

Finally, a similar vertical non-uniformity is also observed in the GC cells with calibration coefficients varying from $\sim -2\%$ to $\sim +2\%$ between the bottom and the top for long cells and from $\sim -4\%$ to $\sim +4\%$ for the short cells. The collected charge decreases with the distance between the interaction point and the PMTs due to the LS attenuation length and other light collection effects: different average number of photon reflections before detection, reflection angle, incidence angle on the PMT photocathode, etc. For long GC cells, the non-uniformity along the longitudinal axis is more significant. The difference between central and edge calibration coefficients is (6 - 8)\% due to the energy leakage outside the active volume of the detector and light collection effects.\\

\section{Energy Scale}
\label{sec:Erec}

The light cross-talks between the detector cells, and particularly their variation in time, are difficult to describe with high accuracy in the simulation. Therefore a procedure of energy reconstruction has been developed, which corrects to first order for all light collection effects. This way, data can be accurately compared with a single reference MC. After presenting the formalism, this section describes how the light leak and calibration coefficients discussed above are fine-tuned to anchor the detector calibration on the energy of the $^{54}$Mn source. The systematic uncertainties of the energy scale based on the observed time stability of the mean detector response, and the residual cell-wise differences between experimental and simulated responses are discussed.

\subsection{Formalism of energy reconstruction}
\label{subsec:Formal_Erec}

Considering an energy deposition $E_i$ in one single cell $i$, the total charge $Q_i$ collected by the 4 PMTs of this cell can be written as
\begin{equation}
  \label{eq:energy_to_charge_conversion}
  Q_i = E_i C_i,
\end{equation}
\noindent{}where $C_i$ is the calibration coefficient of the cell $i.e.$ the product of the number of photons produced per deposited energy and the light collection efficiency of the cell. Because of the light cross-talk, a neighboring cell $j$ will collect the charge $Q_j=L_{ij}Q_i$ according to the definition of the light cross-talk coefficients in Eq.\ref{eq:crosstalk_coefficient}. With Eq.\ref{eq:energy_to_charge_conversion}:
\begin{equation}
  \label{eq:leaks_to_charge_conversion}
  Q_j = E_i C_i L_{ij}.
\end{equation}

The description of any energy deposition distribution among the 10 detector cells can be generalized to
\begin{equation}
\label{eq:charge_general_expression}
  Q_j = \sum_{i = 0}^{9} E_i M_{ij},
\end{equation}
\noindent{}where the matrix notation $M_{ij} \doteq C_iL_{ij}$ has been introduced and by definition $L_{ii}=1$. The $i$ index runs on all the optical volumes of the detector: the front GC, the 6 TG cells, the back GC and the 2 long GC cells. Then by inverting the matrix $M$, the vector of deposited energies can be reconstructed from the vector of detected charges
\begin{equation}
\label{eq:energy_reconstruction}
\left( \begin{matrix} 
  E_0 \\
  E_1 \\
  \vdots \\
  E_9
 \end{matrix}\right)
=
M^{-1}
\left( \begin{matrix} 
  Q_0 \\
  Q_1 \\
  \vdots \\
  Q_9

 \end{matrix}\right).
\end{equation}

\subsection{Fine tuning of the coefficients of energy reconstruction with a $^{54}$Mn source}
\label{subsec:Calib_Mn}


The fine tuning of the energy reconstruction is an iterative procedure that matches the most probable values of reconstructed energies $E^{\text{rec}}_{\text{Data}}$ and $E^{\text{rec}}_{\text{MC}}$ to a common reference $E^{\text{ref}}$. Only events with most of the energy being deposited in a chosen cell $i$ are selected ($E^{\text{rec}}_{i\neq j}<$ 100~keV)\footnote{During phase-I only the TG cells 1, 4 and 6 were equipped with calibration tubes, therefore the chosen cell could be different from, but at least contiguous to, the source cell.}. Then, the reference energy is built from the distribution of the "true" deposited energy of the MC, strongly peaking at the nominal $\gamma$-ray energy. For a consistent comparison between $E^{\text{rec}}_{\text{Data}}$ and $E^{\text{rec}}_{\text{MC}}$, the distribution of true energy is convolved with the detector resolution and the resulting most probable value is taken as $E^{\text{ref}}$. It was verified that changing the resolution function from a simple Gaussian with fixed sigma to the most realistic detector response had virtually no impact on the determined $E^{\text{ref}}$. A 0.2\% uncertainty covers all the observed variations.

\begin{figure}[h!]
	\centering
	\subfigure[]{\includegraphics[width=0.48\linewidth]{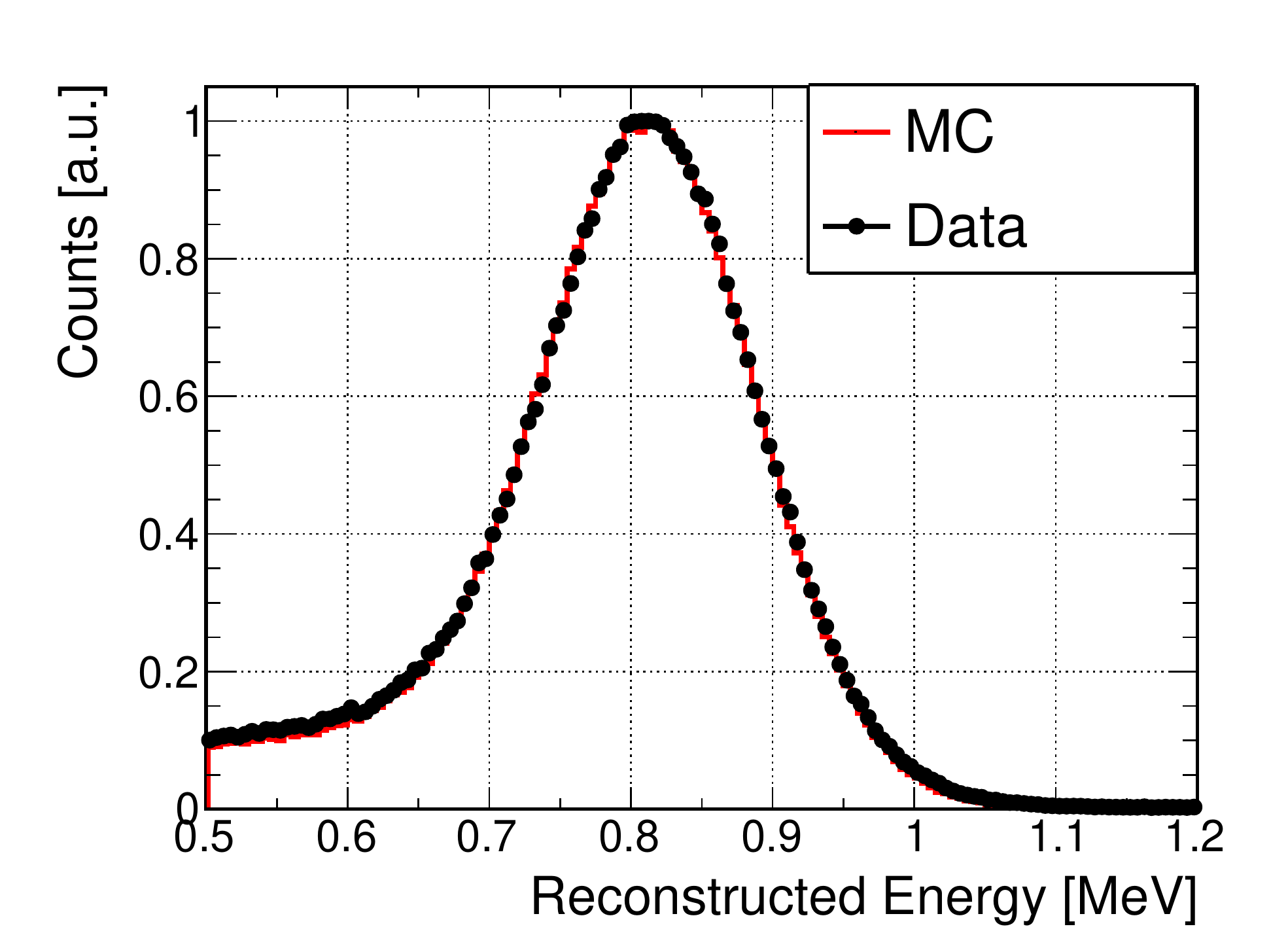}}\quad
	\subfigure[]{\includegraphics[width=0.48\linewidth]{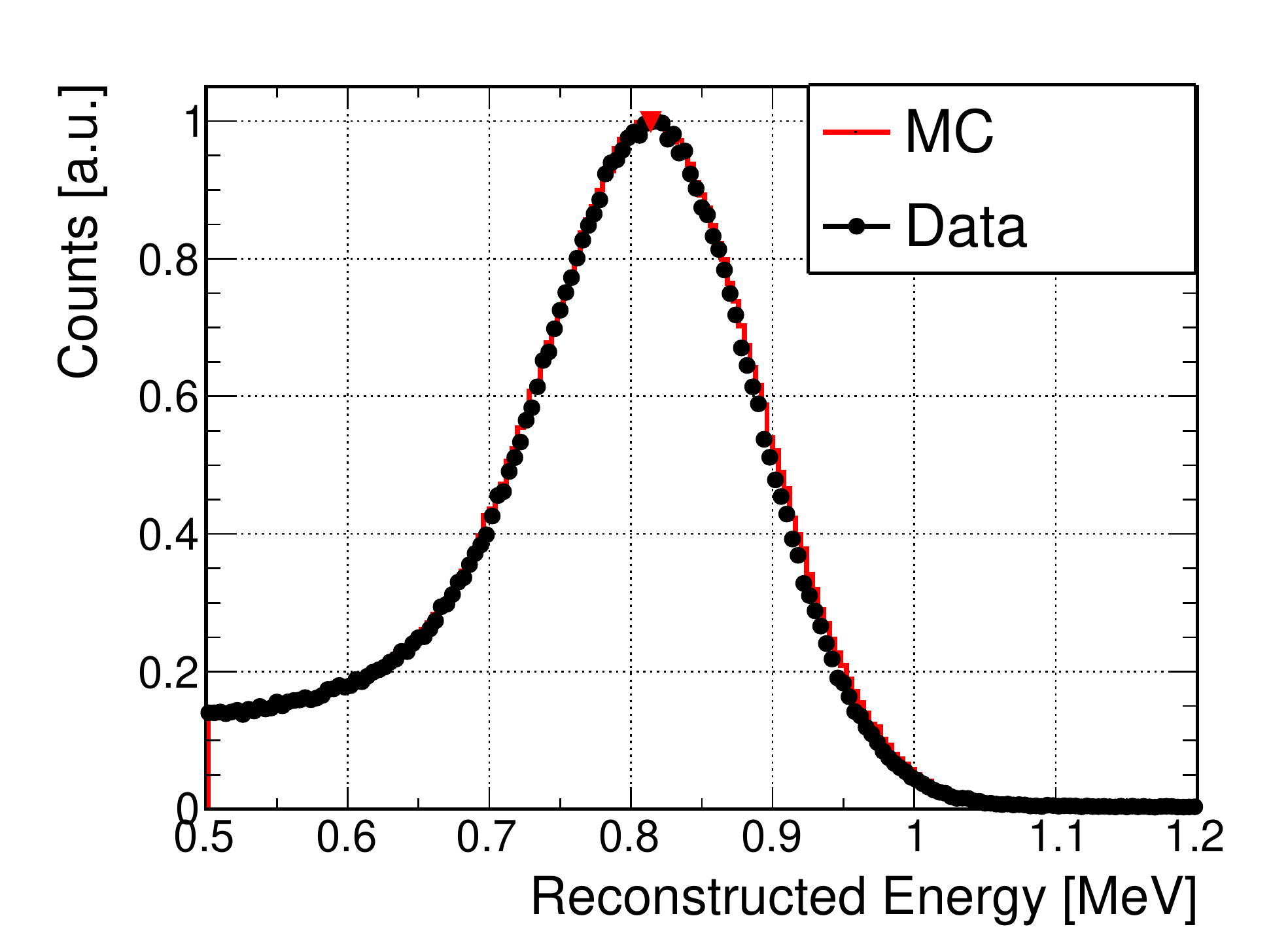}}\\
	\caption{Distribution of reconstructed energies after fine-tuning of the light collection coefficients for a $^{54}$Mn source in the center of cell 6 (a) and for the average of the 5 vertical positions illustrated in Figure~\ref{fig:ZDependence} (b).}
\label{fig:Erec_match}
\end{figure}

Starting from the light cross-talk and calibration coefficients presented in Section \ref{subsec:LightCrossTalk} and \ref{subsec:cell-by-cell-calib}, $E^{\text{rec}}_i$ is first compared with $E^{\text{ref}}_i$ for each chosen cell $i$. The matrix formalism clearly shows that the residual difference is driven by the bias on $C_i$, which is corrected accordingly for all cells. Then the vector of reconstructed energies is computed again, this time including $E^{\text{rec}}_{j\neq i}$. As they correspond to negligible energy deposit, their value is driven by the bias on $L_{ij}$, which is corrected as well. These two steps are repeated in an iterative process until a convergence is reached for all coefficients. In practice, the convergence is reached after a few iterations and the correction to the initial coefficients stays within few percents. The source configurations are either the mid-height point or the average of all measured heights to include the volume effects in the mean calibration coefficients. Figure~\ref{fig:Erec_match} illustrates the excellent agreement obtained between $E^{\text{rec}}_{\text{Data}}$ and $E^{\text{rec}}_{\text{MC}}$ for both cases. Fitting the full energy peak with a gaussian function yields an energy resolution ($\sigma$/E) of 9.0\% with the source at the center, and 9.4\% when averaging all measured heights.

Based on the 41 calibration runs taken during the first phase of \STEREO{}, the residual relative difference between $E^{\text{rec}}_{\text{Data}}$, $E^{\text{rec}}_{\text{MC}}$ and $E^{\text{ref}}$ is centered at 0 with a standard deviation of $0.2\%$ for TG cells and $0.5\%$ for GC cells.

\subsection{Detector stability}
\label{subsec:Stability}

In order to check the quality of the energy reconstruction when applied on events at higher energy and distributed across the whole detector volume, as the neutrino signal is known to be, nuclear spallation events induced by atmospheric muons nearby the detector are selected. Among other particles, spallation neutrons are produced, some of which enter the detector and are captured by hydrogen or gadolinium nuclei, generating a 2.22~MeV single $\gamma$-ray or a $\gamma$-cascade of total energy around 8 MeV, respectively, distributed in all cells. The high multiplicity of the Gd $\gamma$-cascade implies a high probability of the event energy to be shared among several cells. Thus, the total energy and the topology of energy deposits of these events both significantly differ from the $^{54}$Mn calibrations and constitute a stringent test of the energy reconstruction.

\begin{figure}[ht]
\centering
\subfigure[]{\includegraphics[width=0.48\linewidth]{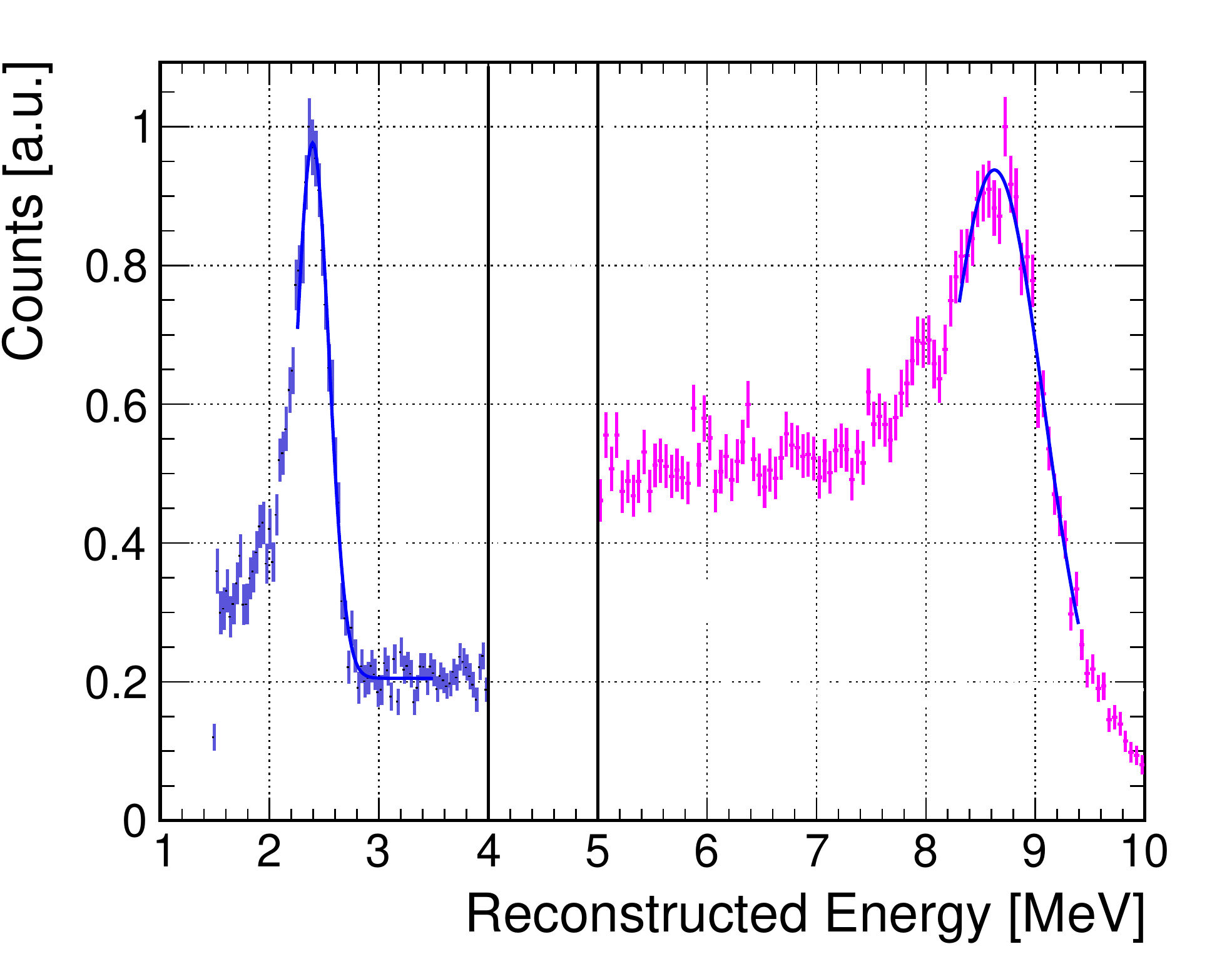}}\quad
\subfigure[]{\includegraphics[width=0.48\linewidth]{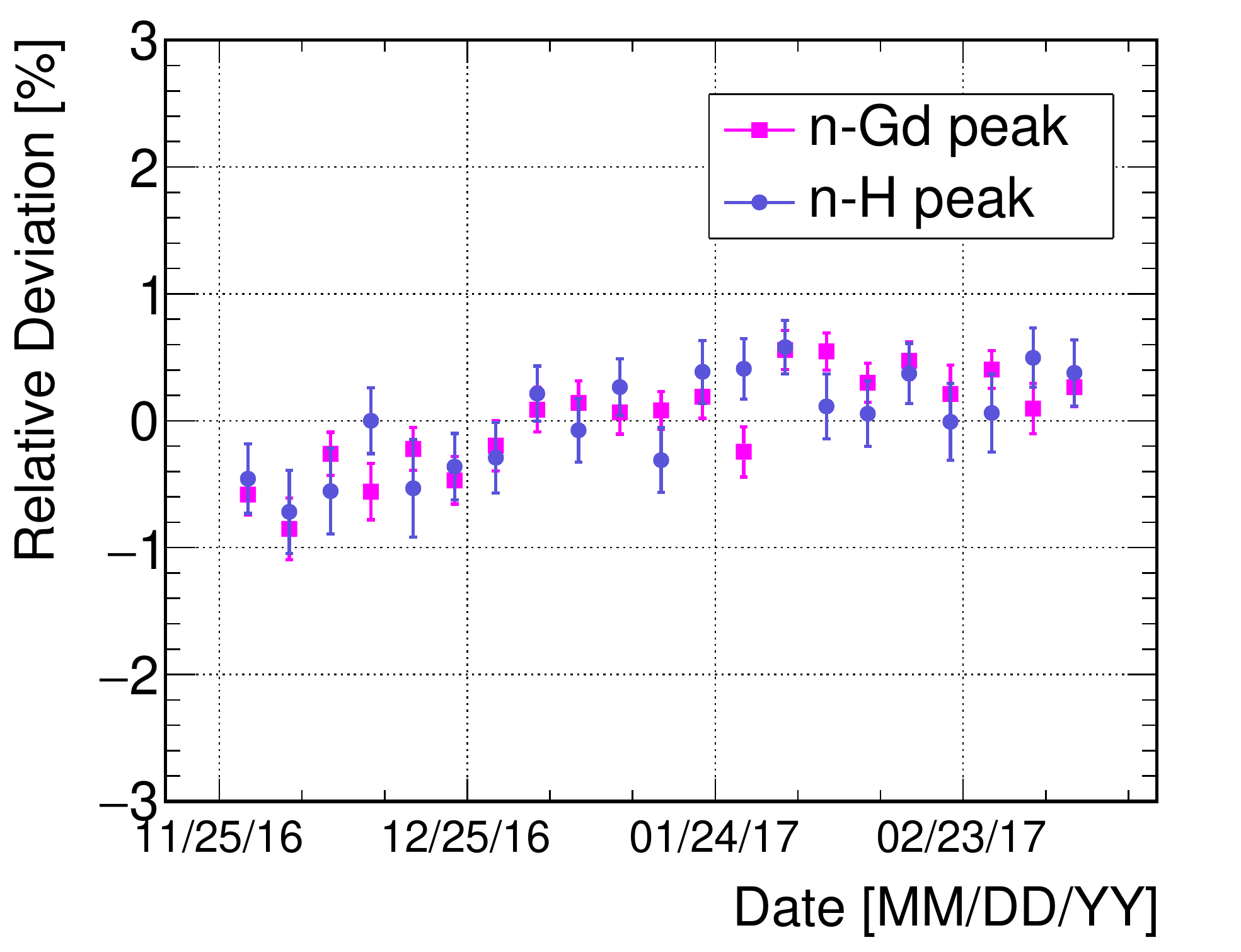}}\\

\caption{(a) Typical example of reconstructed energy for a selection of n-H and n-Gd events with main energy deposition in the TG. (b) Time evolution of the fitted peak positions around their mean value. Each point represents a 5 days period.}
\label{fig:nH_nGd_Stability}
\end{figure}

Typical n-capture peaks are illustrated in Figure~\ref{fig:nH_nGd_Stability} with an example of a fit of their most probable value, and the stability of the peak positions observed in the TG volume for 21 periods of 5 days over the period December 2016 and March 2017. Despite the significant variation of the light cross-talk occurring in the detector during December 2016 (see Figure~\ref{fig:LLEvolution}), the reconstructed peaks are stable at the sub-percent level, with remarkable agreement between the residual common trends of the n-H and n-Gd events. This trend is not corrected for, but instead a 0.35\% relative uncertainty on the energy scale is added which covers the time evolution and allows to use a constant reconstructed energy. 

As clearly seen on Figure~\ref{fig:nH_nGd_Stability}.(a) the mean reconstructed energies of the n-H and n-Gd peaks, 2.38 and 8.63 MeV, are higher than the nominal values of 2.22 and 8.04 MeV\footnote{Estimated by averaging the sum of the gammas emitted by neutron captures in $^{157}$Gd and $^{155}$Gd, weighted by their abundance and cross section.} respectively. This deviation is expected from the quenching effect of the LS discussed in the next section.  



\subsection{Quenching}
\label{subsec:Quenching}

The reference calibration at 0.835 MeV described above provides accurate light cross-talk coefficients, which remain valid for the whole energy range of the neutrino signal since they are ratios of charges (Eqs.\ref{eq:energy_to_charge_conversion} and \ref{eq:leaks_to_charge_conversion}). However, the calibration coefficients suffer from a sizable non-linearity due to the quenching of the LS responses. This behavior can be described by the Birks law \cite{birks}:
\begin{equation}
\frac{\text {d}L}{\text {d}x} = S \frac{\frac{\text {d}E}{\text {d}x}}{1+k_\text {B}\frac{\text {d}E}{\text {d}x}}
\label{eq:birks}
\end{equation}
where $\text {d}L/\text {d}x$ is the emitted energy in the form of light per unit path,  $S$ is the scintillation efficiency and $\text {d}E/\text {d}x$ is the energy deposit per unit path. 
The Birks coefficient $k_\text{B}$ expresses the fact that the LS becomes less efficient at converting deposited energy into light when d$E$/d$x$ becomes large. Thus, the effect is more pronounced for heavier or low energy particles. A consistent analysis requires data and MC to show the same quenching effect. To this end, the value of $k_\text{B}$ is tuned in the simulation to match the quenching observed in data.

\begin{figure}[h!]
	\centering
	\includegraphics[width=0.75\textwidth]{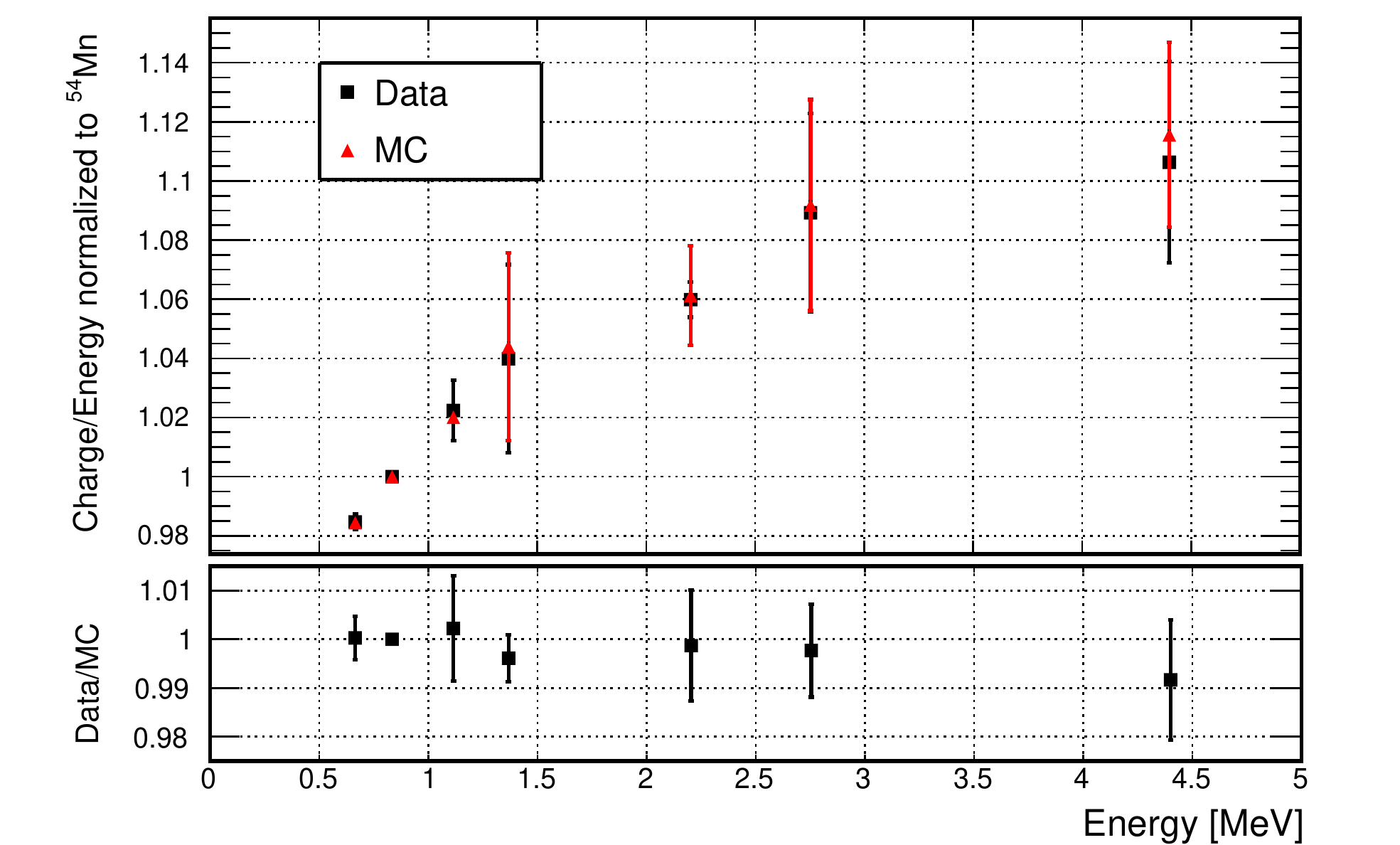}
	\caption{Quenching effect in the TG liquid scintillator using different $\gamma$-ray sources at the half of the cell height. Data and MC points are normalized to the $^{54}$Mn anchor point and $k^{\text{ref}}_\text{B}=0.096$ mm/MeV is used in the simulation.}
	\label{fig:quenching}
\end{figure}	

 The experimental quenching effect is determined using a set of radioactive sources (see Table~\ref{tab:set_sources}) placed inside cell 6 at the half of the cell height. These sources emit single or multiple $\gamma$-rays but the analysis always isolates a unique $\gamma$-ray energy.  
Figure~\ref{fig:quenching} shows the most probable collected charge per MeV, relative to the $^{54}$Mn anchor point, as a function of $E^{\text{ref}}$, the most probable deposited energy predicted by the simulation. The error bars of the top plot represent the systematic uncertainties derived from the different studies of the quenching curve. More specifically, they include the discrepancies obtained for each point from varying the range of the fits and the cell where the full energy of the event has been deposited. Uncertainties of the bottom plot are lower since some systematic uncertainties are common to data and MC and cancel out for their ratio. As expected low energy $\gamma$-rays appear more quenched. The Birks coefficient in the MC was then varied until a good match with the experimental curve was obtained, defining an optimal $k^{\text{ref}}_\text{B}= 0.096 \pm 0.007$ mm/MeV. The bottom curve of Figure~\ref{fig:quenching} shows that for this optimal $k_\text{B}$, data and MC agree at the 1\% level for all energies indicating an accurately tuned $k_\text{B}$.

\subsection{Systematic uncertainty of the energy scale}
\label{subsec:SystEscale}
Based on the above studies, Table \ref{tab:EscaleSyst} summarizes all contributions to the systematic uncertainty of the energy scale. In the oscillation analysis, they will be propagated as uncertainties of a global calibration factor, i.e. fully correlated among all energy bins. The contribution from $E^{\text{ref}}$ is directly taken from Section \ref{subsec:Calib_Mn}. Considering positron events, the 0.2\% uncertainty of the TG cell calibration anchoring on the $^{54}$Mn energy is increased to 0.3\% to take into account the potential energy leak of a 511~keV $\gamma$-ray toward the GC (with 0.5\% uncertainty). From Section \ref{subsec:Stability}, the detector response is known to vary in time at the sub-percent level. As for the other sources of uncertainties, the residual trend is not corrected for. Instead, the uncertainty corresponds to the observed variations of the detector response. As the time stability is measured on the mean response of all the TG cells, it is treated as fully correlated among all cells while all other contributions are uncorrelated.

\begin{table}[h]
	\centering
	\begin{tabular}{ |r| c |  c| }
	\hline
 	Source  & Cell-to-cell &  Cell-to-cell  \\
   	& correlated & uncorrelated  \\ \hline\hline
 	$E^{\text{ref}}$  & - & 0.20\% \\
 	$^{54}$Mn anchor  & - & 0.30\% \\
 	Time stability (n-H)  & 0.35\% & - \\
 	Cell-wise Data-MC comparison &  & \\ 
	(from calib. sources + n-H)  & - & 1.00\% \\  \hline \hline
 	\bf{Total} & \bf{0.35\%} & \bf{1.06\%} \\
	\hline
	\end{tabular}
	\caption{Summary of the uncertainties of the energy scale.\label{tab:EscaleSyst}}
\end{table}

\begin{figure}[h!]
	\centering
	\includegraphics[width=0.65\textwidth]{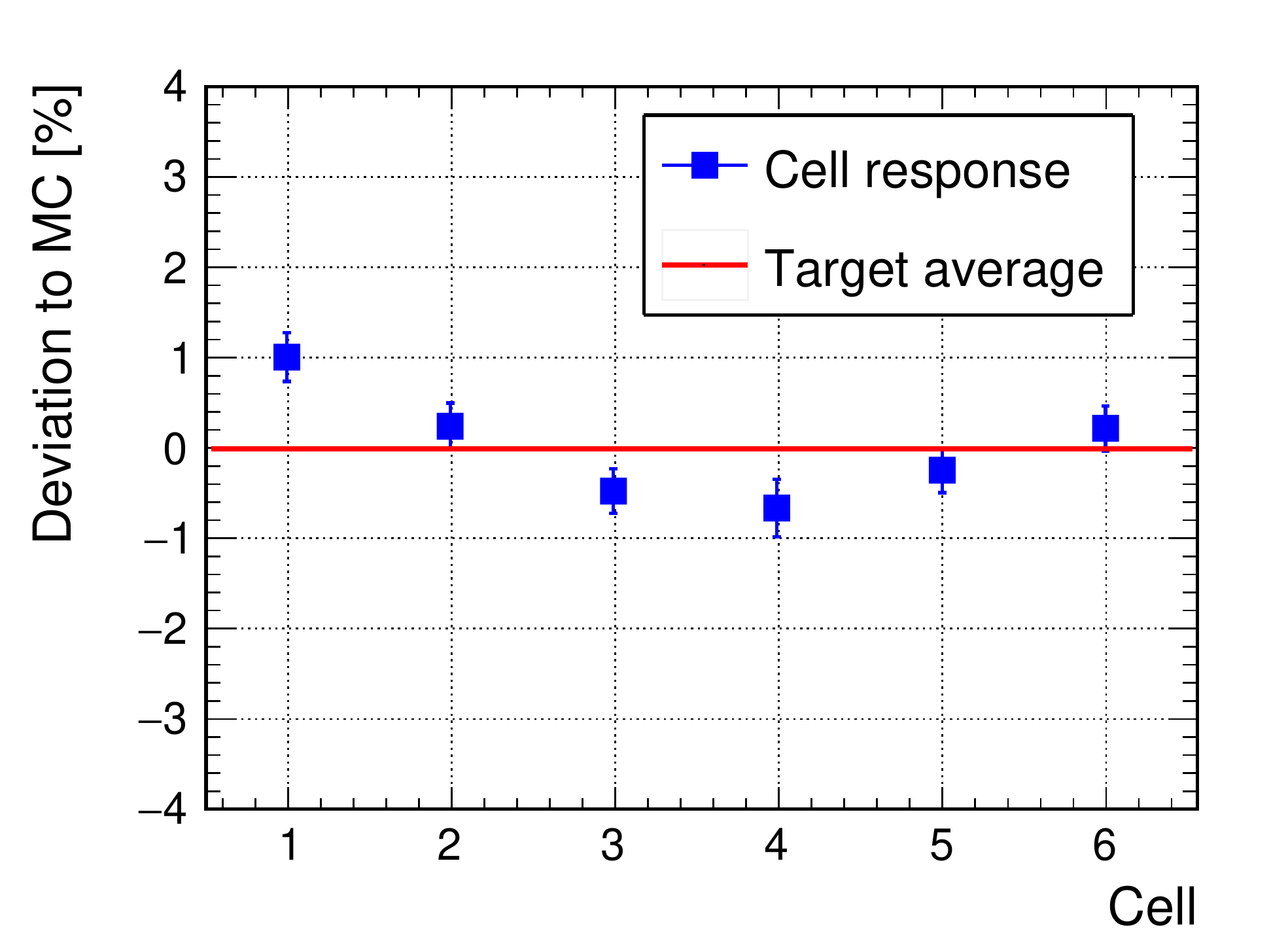}
	\caption{Cell-wise deviation of the experimental most probable energy reconstructed for the n-H events, relative to the simulated energy}
	\label{fig:Cell-wise-Erec}
\end{figure}

The last contribution in Table \ref{tab:EscaleSyst} is the cell-wise relative difference between data and MC reconstructed energies. The residual plot at the bottom of Figure~\ref{fig:quenching} gives a first hint of this uncertainty as data and MC are shown to follow the same energy curve at the 1\% level for all radioactive sources.
In order to test the energy reconstruction with a complementary class of events, the 2.22~MeV $\gamma$-rays emitted after the n-H captures is used. From the GEANT4 simulation of muons interacting nearby the detector, a large majority of the neutron vertices contributing to the selected n-H captures comes from spallation reactions in the lead layer underneath the detector vessel. Thus, for the specific data-MC comparison illustrated in Figure~\ref{fig:Cell-wise-Erec}, the procedure of the energy reconstruction was anchored to the detector response with the $^{54}$Mn source located close to the bottom of the cells rather than averaged along their full vertical extension. 
A similar 1\% level agreement is observed, which is taken as the contribution to the uncertainty of the energy scale.

\begin{figure}[h!]
	\centering
	\includegraphics[width=0.65\textwidth]{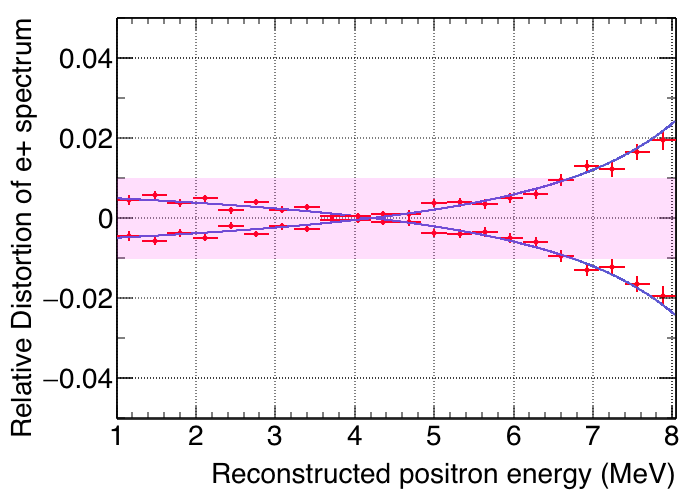}
	\caption{Ratio of distorted to nominal positron spectra when the $k_\text{B}$ is varied by $\pm 0.007$ mm/MeV in the simulation. The shaded area illustrates the limit of $\pm 1\%.$ distortions.}
	\label{fig:Delta_kB}
\end{figure}

Lastly, the impact of the uncertainty of $k^{\text{ref}}_B$ is determined by comparing the nominal positron spectrum and the positron spectra simulated with $k_\text{B} = k^{\text{ref}}_\text{B} \pm$ 0.007 mm/MeV (1 $\sigma$). The induced relative distortions are shown in Figure~\ref{fig:Delta_kB}. They are taken as an energy dependent uncertainty of the simulated spectra, fully correlated among all cells and all energy bins. The amplitude of the distortion stays below 1\% up to 7 MeV, beyond which energy only a very low neutrino rate is expected. This systematic is thus expected to have very little impact on the oscillation analysis.

\section{Conclusions}
The \STEREO{} detector has been successfully installed at the ILL, after three years of design and construction, and started data taking in November 2016. Composed of a 6-cells target filled with gadolinium-loaded liquid scintillator, the \STEREO{} detector is able to measure neutrino spectra at 6 different distances from the reactor core to investigate a hypothetical neutrino oscillation to a new sterile state. During the phase-I, the experiment took 84 days of data with the reactor in operation but also 32 days of data with the reactor stopped, essential for the study of the background induced by cosmic rays. The measured accidental background rates validate the effectiveness of the different shielding layers which have been added around the detector. As the oscillation analysis relies on the comparison of the cells energy spectra, extensive studies of the detector response have been performed. The large photostatistics of the detector, up to 260 PE/MeV in one cell at 1~MeV, meets initial expectations and extensive calibration and monitoring procedures have demonstrated that the energy scale is controlled at the 1\% level. A dedicated energy reconstruction method corrects for all detector drifts and allows to reach a sub-percent stability. 

\begin{figure}[h!]
	\centering
	\includegraphics[width=0.65\textwidth]{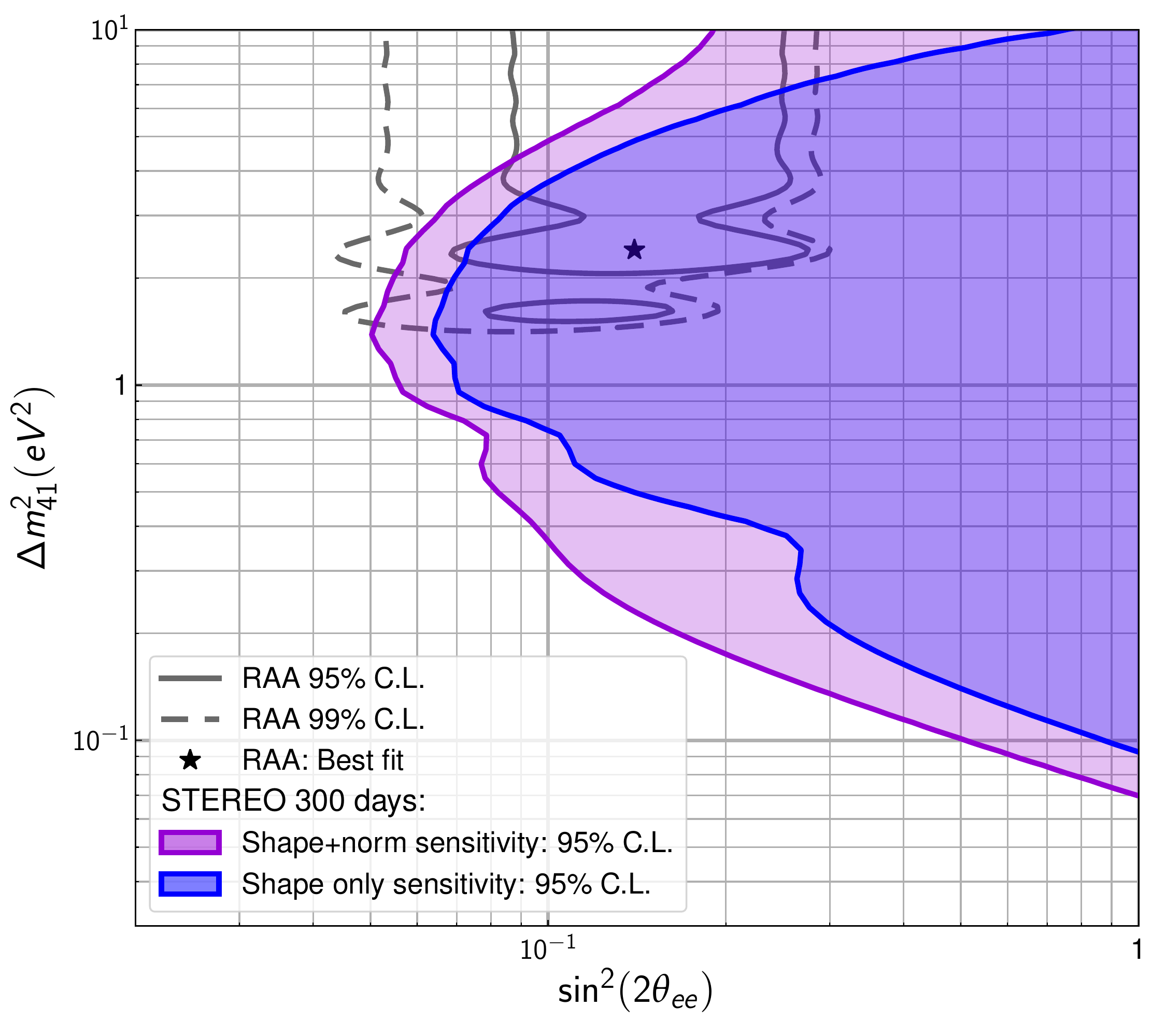}
	\caption{Expected sensitivity to the null oscillation hypothesis of the \STEREO{} experiment for 300 days of reactor-on data, assuming the detector performances reported in this paper and a signal to background ratio equal to 1, estimated within our current selection cuts. The RAA contours and the best fit point are taken from \cite{Mention}.}
	\label{fig:Sensitivity}
\end{figure}

These performances demonstrate that \STEREO{} has reached the required control of all detection uncertainties to address the existence of the sterile neutrino at the eV scale. The experiment is currently running in phase-II where it will take data until the end of 2019 for a total of 6 reactor cycles of 50 days each. The expected sensitivity to the null oscillation hypothesis assuming such a statistic and the detector performances reported in this paper is shown in Figure~\ref{fig:Sensitivity}. The \STEREO{} measurement is expected to cover the RAA allowed region at 95\% C.L.

\acknowledgments
This work is funded by the French National Research Agency (ANR) within the project ANR-13-BS05-0007 and the 'Investments for the future' programs P2IO LabEx (ANR-10-LABX-0038) and ENIGMASS LabEx (ANR-11-LABX-0012).
Authors are grateful for the technical and administrative support of the ILL for the installation and operation of the \STEREO{} detector. In particular, we thank S. Armanet, K. Ben Salem, M. Carrier, F. Chanteloup, P. Cogo, M. David,  L. Didier, O. Equilbec, A. Eyssard, R. Gandeli, Y. Gibert, C. Gillart, B. Glyn-Jones, B. Jarry, P. Lacombe, J. Lauga, Y. Le Goc, P. Lachaume, J. Locatelli, N. Marconnet, T. Mazzilli, P. Mutti, I. Perbet, L. Pereira Bahia, F. Rencurel, G. Rochex, M. Samuel, I. Sives, A. Saragaglia, F. Thomas as representatives for many others.

\end{document}